\documentclass[aps,twoside,showpacs,showkeys,superscriptaddress,11pt]{revtex4}

\usepackage{graphicx,epic,eepic,epsfig,amsmath,latexsym,amssymb,verbatim}

\usepackage{theorem}
\newtheorem{definition}{Definition}[section]

\newtheorem{lemma}[definition]{Lemma}

\newtheorem{theorem}[definition]{Theorem}
\newtheorem{corollary}[definition]{Corollary}

\def\squareforqed{\hbox{\rlap{$\sqcap$}$\sqcup$}}
\def\qed{\ifmmode\squareforqed\else{\unskip\nobreak\hfil
\penalty50\hskip1em\null\nobreak\hfil\squareforqed
\parfillskip=0pt\finalhyphendemerits=0\endgraf}\fi}
\def\endenv{\ifmmode\;\else{\unskip\nobreak\hfil
\penalty50\hskip1em\null\nobreak\hfil\;
\parfillskip=0pt\finalhyphendemerits=0\endgraf}\fi}
\newenvironment{proof}{\noindent \textbf{{Proof~} }}{\qed}
\newenvironment{remark}{\noindent \textbf{{Remark~}}}{\qed}

\mathchardef\ordinarycolon\mathcode`\:
\mathcode`\:=\string"8000
\def\vcentcolon{\mathrel{\mathop\ordinarycolon}}
\begingroup \catcode`\:=\active
  \lowercase{\endgroup
  \let :\vcentcolon
  }

\newcommand{\nc}{\newcommand}
\nc{\rnc}{\renewcommand}
\nc{\beq}{\begin{equation}}
\nc{\eeq}{{\end{equation}}}
\nc{\bea}{\begin{eqnarray}}
\nc{\eea}{\end{eqnarray}}
\nc{\beqa}{\begin{eqnarray}}
\nc{\eeqa}{\end{eqnarray}}
\nc{\lbar}[1]{\overline{#1}}
\nc{\bra}[1]{\langle#1|}
\nc{\ket}[1]{|#1\rangle}
\nc{\ketbra}[2]{|#1\rangle\!\langle#2|}
\nc{\braket}[2]{\langle#1|#2\rangle}
\nc{\proj}[1]{| #1\rangle\!\langle #1 |}
\nc{\avg}[1]{\langle#1\rangle}
\rnc{\max}{\operatorname{max}}
\nc{\Rank}{\operatorname{Rank}}
\nc{\smfrac}[2]{\mbox{$\frac{#1}{#2}$}}
\nc{\Tr}{\operatorname{Tr}}
\nc{\ox}{\otimes}
\nc{\dg}{\dagger}
\nc{\dn}{\downarrow}
\nc{\cA}{{\cal A}}
\nc{\cB}{{\cal B}}
\nc{\cC}{{\cal C}}
\nc{\cD}{{\cal D}}
\nc{\cE}{{\cal E}}
\nc{\cF}{{\cal F}}
\nc{\cG}{{\cal G}}
\nc{\cH}{{\cal H}}
\nc{\cI}{{\cal I}}
\nc{\cJ}{{\cal J}}
\nc{\cK}{{\cal K}}
\nc{\cL}{{\cal L}}
\nc{\cM}{{\cal M}}
\nc{\cN}{{\cal N}}
\nc{\cO}{{\cal O}}
\nc{\cP}{{\cal P}}
\nc{\cR}{{\cal R}}
\nc{\cS}{{\cal S}}
\nc{\cT}{{\cal T}}
\nc{\rU}{{\cal U}}
\nc{\cX}{{\cal X}}
\nc{\cZ}{{\cal Z}}
\nc{\csupp}{{\operatorname{csupp}}}
\nc{\qsupp}{{\operatorname{qsupp}}}
\nc{\var}{\operatorname{var}}
\nc{\rar}{\rightarrow}
\nc{\lrar}{\longrightarrow}
\nc{\poly}{\operatorname{poly}}
\nc{\polylog}{\operatorname{polylog}}
\nc{\1}{\openone}

\def\>{\rangle}
\def\<{\langle}

\def\a{\alpha}
\def\b{\beta}
\def\g{\gamma}
\def\d{\delta}
\def\e{\epsilon}

\def\h{\eta}

\def\l{\lambda}

\def\r{\rho}
\def\s{\sigma}

\def\ph{\varphi}

\def\ps{\psi}

\def\G{\Gamma}

\nc{\glneq}{{\raisebox{0.6ex}{$>$}  \hspace*{-1.8ex} \raisebox{-0.6ex}{$<$}}}
\nc{\gleq}{{\raisebox{0.6ex}{$\geq$}\hspace*{-1.8ex} \raisebox{-0.6ex}{$\leq$}}}

\nc{\RR}{{{\mathbb R}}}
\nc{\CC}{{{\mathbb C}}}
\nc{\FF}{{{\mathbb F}}}
\nc{\NN}{{{\mathbb N}}}
\nc{\ZZ}{{{\mathbb Z}}}
\nc{\PP}{{{\mathbb P}}}
\nc{\QQ}{{{\mathbb Q}}}
\nc{\UU}{{{\mathbb U}}}
\nc{\WW}{{{\mathbb W}}}
\nc{\EE}{{{\mathbb E}}}
\rnc{\SS}{{{\mathbb S}}}
\nc{\id}{{\operatorname{id}}}

\nc{\vholder}[1]{\rule{0pt}{#1}}

\nc{\ob}[1]{#1}

\nc{\eq}[1]{Eq.~(\ref{eq:#1})}
\nc{\eqs}[2]{Eqs.~(\ref{eq:#1}) and (\ref{eq:#2})}

\nc{\eqn}[1]{Eq.~(\ref{eqn:#1})}
\nc{\eqns}[2]{Eqs.~(\ref{eqn:#1}) and (\ref{eqn:#2})}

\begin{document}

\title{{\Large Aspects of generic entanglement}}

\author{Patrick Hayden}
\email{patrick@cs.mcgill.ca}
\affiliation{Institute for Quantum Information, Caltech 107--81,
    Pasadena, CA 91125, USA}
\affiliation{Department of Computer Science, McGill University,
    Montreal, Quebec, Canada H3A 2A7}
\author{Debbie W. Leung}
\email{wcleung@cs.caltech.edu}
\affiliation{Institute for Quantum Information, Caltech 107--81,
    Pasadena, CA 91125, USA}
\author{Andreas Winter}
\email{a.j.winter@bris.ac.uk}
\affiliation{Department of Mathematics, University of Bristol,
Bristol BS8 1TW, United Kingdom}

\date{10th June 2004}

\begin{abstract}
We study entanglement and other correlation properties of random
states in high-dimensional bipartite systems. These correlations are
quantified by parameters that are subject to the ``concentration of
measure'' phenomenon, meaning that on a large-probability set these
parameters are close to their expectation.
For the entropy of entanglement, this has the counterintuitive consequence
that there exist large subspaces in which all pure states are close to
maximally entangled.
This, in turn, implies the existence of mixed states with entanglement
of formation near that of a maximally entangled state, but with
negligible quantum mutual information and, therefore, negligible
distillable entanglement, secret key, and common randomness.
It also implies a very strong locking effect for the entanglement of
formation: its value can jump from maximal to near zero by tracing over
a number of qubits negligible compared to the size of total system.
Furthermore, such properties are generic.
Similar phenomena are observed for random multiparty states, leading
us to speculate on the possibility that the theory of entanglement is
much simplified when restricted to asymptotically generic states.
Further consequences of our results include a complete
derandomization of the protocol for universal superdense coding of
quantum states.  \par

\end{abstract}



\maketitle

\parskip .75ex

\section{Introduction} \label{sec:intro}

The subject of correlations between quantum systems can be
bewildering.
Beyond the simplest example, manipulation of pure
bipartite states, very little is known.
Exotic examples have implied that the rules governing interconversion
of quantum states are often counterintuitive.
The complexity of the subject is also manifested in the difficulties
one encounters when attempting to quantify entanglement.
Even in the bipartite, {\em asymptotic} case, there are many different
mixed-state ``entanglement measures,'' most of which are poorly
understood, both individually and in relation to each other;
see~\cite{BDSW96,VPRK97,HHT00,MH01,DHO01,CW03} and references therein.


One of the most striking features of asymptotic entanglement
manipulations is irreversibility.  Even in the limit of large number
of copies, some states cost more EPR pairs to create than can be
distilled from them.  The corresponding entanglement measures, known as
the entanglement cost ($E_c$) \cite{HHT00} and the entanglement of
distillation ($E_d$) \cite{BDSW96}, are therefore different.
In particular, for
some ``bound entangled'' states \cite{HHH98}, it has been shown that
$E_d$ is zero while $E_c$ is not~\cite{VC01,VDC02}.

Another intriguing issue in the study of entanglement is whether the
entanglement cost of a state is equal to a much simpler measure, the
entanglement of formation ($E_f$) \cite{BDSW96}.  If equality holds,
the study of entanglement can be simplified significantly, while
inequality implies the advantage of more collective strategies in the
asymptotic preparation of quantum states.  This problem has recently
been connected to other important additivity conjectures in quantum
information theory. (See, for example,~\cite{S04}.)

While the general theory of entanglement appears to be very complicated, a much
simplified theory may exist for generic quantum states if, in some
appropriate regime, most states behave similarly.
In particular, irreversibility has only been demonstrated for some
carefully constructed states and it is natural to wonder whether it is
the exception or the rule.
Here, we investigate possible simplifications of the entanglement
properties of quantum states in large systems.

Considerable effort has been devoted to understanding the average
properties of quantum states.
%
%
For example, the expected entropy \cite{LP88,P93,FK94,S-R95,S96} and purity
\cite{L78,ZZF00,ZS01} of reduced states for random pure quantum states
have been calculated.
In the case of mixed states, various distributions have been
proposed (see \cite{ZS01} and references therein) and the likelihood
of separable (i.e., $E_f=0$) and bound entangled states have been
studied \cite{ZHSL98,Z99,KZM02,S03}.

The present paper is a further step in the direction of a simplified
theory of entanglement for generic states.
We draw random pure states from the uniform (unitarily invariant)
distribution and mixed states by tracing over part of a random
pure state on an extended system. (Note that the induced distribution
depends on the dimension of the system that was traced out.)  We find
that random pure states are extremely likely to have near-maximal
entanglement, in fact, so likely that, with high probability, a random
subspace of dimension close to the total dimension contains only
near-maximally entangled states.  These findings imply that random
mixed states of up to almost full rank can have entanglement of
formation close to maximal; at the same time distillable entanglement,
secret key and common randomness can all be bounded by much smaller
quantities.  In fact,
for a wide range of parameters,
%
%
these random mixed states will not be one-copy pseudo-distillable, and
will have arbitrarily small one-way distillable entanglement, secret
key, and common randomness.  Thus, near-maximal irreversibility is
generic, unless our states turn out to be counterexamples to the
additivity conjecture with near-maximal violation.

By building on the results for the bipartite case, we can make
similar high-probability statements about many properties
of random multiparty states as well. We find, for example, that
a typical such state has near-maximal distillable entanglement between
any two parties, provided the other parties are allowed to participate
in the distillation protocol.

Finally, the existence of large subspaces containing only
near-maximally entangled states has applications to the study of
quantum communication, not just the study of correlations. In
particular, it implies that the protocol for superdense coding of
arbitrary $2l$-qubit states using $l$ ebits and $l+o(l)$ qubits of
communication \cite{HHL03} can be completely derandomized. The
original construction consumed $l+o(l)$ shared random bits in addition
to the other resources.

%
\noindent
{\bf Guide to the paper}\quad In Section
\ref{sec:distributions}, we introduce the pure and mixed state
distributions we will be investigating. We discuss our basic
techniques in Section \ref{sec:concentration}.
These consist of an elementary discretization
procedure, which ``counts'' the number of points in a geometrical
manifold (of states, subspaces, etc), and explicit inequalities for
the concentration of measure phenomenon in functions on
high-dimensional spheres.
Our main result, proving that random subspaces are likely to
contain only near-maximally
entangled states, appears in Section \ref{sec:subspace}. We then
study various entanglement, secret key and other correlation
quantities in Section \ref{sec:measures}.  Some preliminary results on
generic multiparty entanglement appear in Section
\ref{sec:multiparty}. Finally, we show how our results derandomize
superdense coding in Section \ref{sec:superdense}.

\noindent
{\bf Notation}\quad
We use the following conventions throughout the paper.  $\log$ and
$\exp$ are always taken base two. Unless otherwise stated, a ``state''
can be pure or mixed.  The symbol for a state (such as $\ph$ or
$\rho$) also denotes its density matrix.  We will make an explicit
indication when referring to a {\em pure} state. The density matrix
$\proj{\ph}$ of the pure state $\ket{\ph}$ will frequently be written
simply as $\ph$. $\cB(\CC^d)$ will be used to denote the set of linear
transformations from $\CC^d$ to itself and $\UU(d) \subset \cB(\CC^d)$
the unitary group on $\CC^d$. $\EE X$ refers to the expectation value of the
random variable $X$ and $m(X)$ a median for $X$.
Quantum systems under consideration will be denoted $A$, $B$, $\ldots$
and are freely associated with their Hilbert spaces, whose (finite)
dimensions are denoted $d_A$, $d_B$, etc.  In a bipartite system, when
speaking of a ``maximally entangled state'', we refer to a pure state
whose nonzero Schmidt coefficients \cite{Nielsen00bk} are all equal to
the inverse of the smaller of the two dimensions.  We use $S(\r) =
-\Tr \r \log \r$ to refer to the von Neumann entropy~\cite{vN27} of a
density matrix $\r$, $S(A:B)_\r = S(\r_A) + S(\r_B) - S(\r_{AB})$
to denote the quantum mutual information of a bipartite
state $\r_{AB}$ and $D(\r\|\s) = \Tr \r (\log \r - \log \s)$ the relative
entropy of the states $\r$ and $\s$~\cite{U62}.
The function $F(\r,\s) = (\Tr\sqrt{\r^{1/2} \s
\r^{1/2}})^2$ is the Uhlmann fidelity \cite{Uhlmann76}, again between two
states $\r$ and $\s$.

\section{Random states and subspaces} \label{sec:distributions}

We are going to consider the state of large bipartite (and later
multipartite) quantum systems under a random selection. We think of
the pure or mixed state as being drawn at random from an ensemble. For
pure states, there is a unique ``uniform'' distribution that is
unitarily invariant. It is induced by the Haar measure~\cite{DP99} on
the unitary group by acting on an arbitrarily chosen generating
vector.

\begin{definition} \label{def:randomvector}
A \emph{random pure state} $\ph$ is any random variable distributed
according to the unitarily invariant probability measure on the pure
states $\cP(A)$ of the system $A$. We formally express this by writing
$\ph\in_R \cP(A)$. (It is frequently convenient to choose a vector
representative in $A$ for the state $\ph$. When doing so, we will indicate
this by using the notation $\ket{\ph}$.)
\end{definition}

Similarly, there is also a unique, uniform distribution for
subspaces that is unitarily invariant.

\begin{definition} \label{def:randomspace}
A \emph{random subspace $S$ of dimension $s$} is any random variable
distributed according to the unitarily invariant measure on the
$s$-dimensional subspaces of $A$, the \emph{Grassmannian}
$\cG_s(A)$ (see, for example,~\cite{Harris92}).
We express this using the notation $S\in_R \cG_s(A)$. Note that $\cG_1(A)$
is naturally isomorphic to $\cP(A)$.
\end{definition}

For mixed states, unitary invariance does not uniquely specify a
probability measure.  Instead, we follow an old proposal to induce
probability measures on mixed states \cite{Braunstein96,Hall98,ZS01}
by partial tracing.

\begin{definition} \label{def:randommixed}
For a system $A$ and an integer $s \geq 1$, consider the distribution
on the mixed states $\cS(A)$ of $A$ induced by the partial trace over
the second factor from the uniform distribution on pure states of $A
\ox \CC^s$.
Any random variable $\rho$ distributed as such will be called a
\emph{rank-$s$ random state}; formally, $\rho \in_R \cS_s(A)$.
Note that the rank of $\rho$ is equal to $\min(s,d_A)$ with
probability $1$.  Also, $\cP(A) = \cS_1(A)$.
\end{definition}

These distributions on states have previously received considerable
interest; so much indeed that the expectation values of several
quantities of interest are known either exactly or to good
approximations.
It is clear that the average of any random rank-$s$ state is the
maximally mixed state, $\frac{1}{d_A}\1$.
We will also make explicit use of the average entropy of a subsystem,
which was conjectured in \cite{P93} and proved in \cite{FK94,S-R95,S96}:

\begin{lemma} \label{lem:entropySubsystem}
Let $\ket{\ph}$ be chosen according to the unitarily invariant measure
on a bipartite system $A \ox B$ with local dimensions $d_A \leq
d_B$, i.e. $\ph \in_R \cP(A \ox B)$. Then
\begin{equation*}
\EE S(\ph_A)
= \frac{1}{\ln 2}\left(
    \sum_{j = d_B + 1}^{d_A d_B} \frac{1}{j} - \frac{d_A - 1}{2d_B}
  \right)
> \log d_A - \smfrac{1}{2} \b \,,
\end{equation*}
where $\b = \frac{1}{\ln 2}\frac{d_A}{d_B}$.
\qed
\end{lemma}
The inequality can be demonstrated by making use of the estimate~\cite{Y91}
\begin{equation*}
\frac{1}{2(d+1)} < \sum_{j=1}^d \frac{1}{j} - \ln d - \g < \frac{1}{2d} \,,
\end{equation*}
where $\g \approx 0.577$ is Euler's constant.

In the following, we will identify the large probability behavior of
functions such as $S(\ph_A)$. It turns out that the probability in
question is often exponentially close to $1$ in some parameter $k$; that is,
for sufficiently large $k$,
the probability is at least $1 - \exp[ - k/ \polylog(k) ]$. We
shall in this case adopt the expression that the behavior in question
\emph{is $k$-likely}. In some cases we won't specify $k$ and will simply
speak of \emph{likely} behavior.

\section{Concentration of measure} \label{sec:concentration}

It is a striking yet elementary fact that the uniform measure on the
$k$-sphere, $\SS^k$, concentrates very strongly about any equator
as $k$ gets large; indeed, any polar cap strictly smaller than a
hemisphere has relative volume exponentially small in $k$.
This simple observation implies a similar result for the value of
any slowly varying function on the sphere, which we can understand as a
random variable induced by the sphere's uniform measure:
namely, it will take values close to the
average except for a set of volume exponentially small in
$k$. Levy's Lemma (Lemma \ref{lem:Levy} below) rigorously formalizes this
idea: ``slow variation'' is encoded as a bound on the Lipschitz
constant of the function (essentially the maximum gradient),
and ``close to the average'' is modelled as a small but finite
deviation. Given only these data, Levy's Lemma gives an explicit
exponential probability bound on the set of ``large deviation''.
Since pure quantum states in $d$ dimensions can be represented as
$2d$-dimensional real unit vectors, the above observations on spheres
ensure that as the dimension of a quantum system becomes large it
comes to make sense to discuss typical behavior of random states, in
the sense that for many properties of interest, almost all quantum
states behave in essentially the same way.

The analysis leading to the various results in this paper will revolve
around the concentration of the spectrum of the reduced density matrix
of a bipartite system when both subsystems are large.  This in turns
implies many important concentration effects.  One example is the
concentration of the entropy of the reduced density matrix (or the
entanglement between the two systems).  Concentration effects for the
maximum and minimum eigenvalues also imply tight bounds on the reduced
density matrix itself and the values of various projections.

Our method of demonstrating generic properties is always to prove that
the opposite is an unlikely event.
We then rewrite the ``bad event'' as a union of ``elementary bad
events'' on a net of states; the cardinality of the net is then
bounded.  In most cases, the cardinality of the net is exponentially
large in the dimension parameter, while the ``elementary bad event''
has an exponentially small probability, due to some measure
concentration.  The probability of the bad event is thus bounded by
the product of these exponentially large and exponentially small
quantities and our goal is to make it (exponentially) less than $1$.

Because we strive for explicit probability and dimension bounds, the
expressions in our theorems and some of the estimates may appear
clumsy at first sight. It is in the nature of the problem (and partly
of our method), however, that the crucial quantities are always
composed of (a) a dimension parameter, which dominates, (b) a
logarithmic factor, (c) a factor quantifying the allowable size of
deviations from the average and (d) an absolute constant.
Our obsession with explicit exponents throughout the paper is needed,
since the exponentially large net size usually allows little
optimization, and everything depends on the achievable strength of
measure concentration.

In the rest of this section, we list a number of basic tools including
concentration effects and net constructions.
Readers who are specifically interested in the correlation properties
of random states could read the statements (without the proofs) of
Levy's Lemma, the concentration of entropy and the existence of small
nets, and move directly to the next section, referring back to the
rest of the tools as necessary rather than trying to absorb them all
beforehand.

\begin{lemma}[Levy's Lemma; see \cite{MS86}, Appendix IV, and \cite{L01}]
\label{lem:Levy}
Let $f : \SS^{k} \rar \RR$ be a function with Lipschitz constant
$\h$ (with respect to the Euclidean norm) and a point $X
\in \SS^{k}$ be chosen uniformly at random.
%
%
Then
\begin{equation*}
\begin{array}{crcrr}
1.~~ &
\Pr\left\{ f(X) - \EE f  ~ \glneq ~ \pm\a \right\}
  &\leq& 2\exp\left( -C_1 (k+1) \a^2 / \h^2 \right) &
~~\mbox{and} \\
2.~~ &
\Pr\left\{ f(X) - m(f) ~ \glneq ~ \pm\a \right\}
  &\leq& \exp\left( -C_2 (k-1) \a^2 / \h^2 \right)
\end{array}
\end{equation*}
for absolute constants $C_i>0$ that may be chosen as
$C_1 = (9\pi^3 \ln 2)^{-1}$ and $C_2 = (2 \pi^2 \ln 2)^{-1}$.
($\EE f$ is the mean value of $f$, $m(f)$ a median for $f$.) \qed
\end{lemma}

We are going to apply Levy's Lemma to the entropy of the reduced state
of a randomly chosen pure state $\ph$ in a bipartite system $A \ox B$,
i.e., $f(\ket{\ph}) = S(\ph_A)$.  Note that $k=2d_Ad_B-1$, and all that
remains is to bound the Lipschitz constant.
\begin{lemma} \label{lem:E-Lipschitz}
The Lipschitz constant $\eta$ of $S(\ph_A)$ is upper bounded by
$\sqrt{8}\log d_A$, for $d_A\geq 3$.
\end{lemma}
\begin{proof}
We first consider the Lipschitz constant of the function $g(\ph) =
H(M(\ph_A))$, where $M$ is any fixed complete von Neumann measurement
and $H$ is the Shannon entropy.  Let $\ket{\ph} = \sum_{jk} \ph_{jk}
\ket{e_j}_A \ket{f_k}_B$ in terms of some orthonormal bases $\{ \ket{e_j}_A
\}$ for $A$ and $\{ \ket{f_k}_B \}$ for $B$.  By unitary invariance, we
may assume that $M_j = \proj{e_j}_A$.  Therefore, if we define
\begin{equation*}
p(j|\ph) = \: {_A \bra{e_j}} \ph_A \ket{e_j}_A = \sum_k |\ph_{jk}|^2 \,,
\end{equation*}
then
\begin{equation*}
g(\ph) = H(M(\ph_A)) = - \sum_j p(j|\ph) \log p(j|\ph) \,.
\end{equation*}
An elementary calculation yields
\begin{eqnarray*}
\h^2
 = \sup_{\braket{\ph}{\ph} \leq 1} \nabla g \cdot \nabla g
&=& \sum_{jk} \frac{4 |\ph_{jk}|^2}{(\ln 2)^2} [ 1 + \ln p(j|\ph) ]^2
     \label{eqn:entropySigma} \\
&\leq& \frac{4}{(\ln 2)^2} [ 1 + \sum_{j} p(j|\ph)
     (\ln p(j|\ph))^2 ] \\
&\leq& \frac{4}{(\ln 2)^2} [ 1 + (\ln d_A)^2 ] \leq 8 (\log d_A)^2,
\end{eqnarray*}
where the second inequality can be shown to hold for $d_A \geq 3$ using
Lagrange multipliers.

Using the above bound, the Lipschitz constant for the von Neumann entropy
$S(\ph_A)$ can be controlled as follows.
Consider any two unit vectors $\ket{\ph}$ and $\ket{\psi}$, and
without loss of generality assume $S(\ph_A) \leq S(\psi_A)$.  If we
choose the measurement $M$ to be along the eigenbasis of $\ph_A$,
$H(M(\ph_A))=S(\ph_A)$ and we have~\cite{vN27}
\begin{equation*}
S(\psi_A) - S(\ph_A) \leq  H(M(\psi_A)) - H(M(\ph_A))
                     \leq  \h \; \| \, \ket{\psi}-\ket{\ph} \, \|_2 \,.
\end{equation*}
%
%
Thus, the Lipschitz constant for $S(\ph_A)$ is bounded by that of
$H(M(\ph_A))$ and we are done.
\end{proof}

\vspace{2ex}

\begin{theorem}[Concentration of entropy] \label{thm:subsystemConvergence}
Let $\ph \in_R \cP(A \ox B)$ be a random state on $A \ox B$, with
$d_B\geq d_A\geq 3$.  Then
\begin{equation*}
\Pr\left\{ S(\ph_A) < \log d_A - \a - \b \right\}
\leq \exp\Big( {-} \frac{(d_Ad_B-1)C_3\a^2}{(\log d_A)^2} \Big) \,,
\end{equation*}
where $\b=\frac{1}{\ln 2}\frac{d_A}{d_B}$ is as in Lemma
\ref{lem:entropySubsystem} and $C_3 = (8 \pi^2 \ln 2)^{-1}$.
\end{theorem}
\begin{proof}
As suggested earlier, we choose $f(\ph) = S(\ph_A)$.
We could use Lemma \ref{lem:Levy}.1 directly but will get better constants
with a bit more work.
We need to relate the median of $f$ to the mean, which is known.
Choose a subset $X$ of the unit ball of $A\ox B$ having relative volume
$1/2$ and such that $\ket{\ph} \in X$ implies that $S(\ph_A) \leq m(f)$.
Then
\begin{eqnarray*}
\log d_A - \smfrac{1}{2}\b \leq \EE f
&=& \int_{X} S(\ph_A) d\ph + \int_{\bar{X}} S(\ph_A) d\ph \\
&\leq& \smfrac{1}{2} m(f) + \smfrac{1}{2} \log d_A.
\end{eqnarray*}
Therefore, $m(f) \geq \log d_A - \b$ and the result follows by
combining Lemmas \ref{lem:Levy}.2 and \ref{lem:E-Lipschitz}.
\end{proof}

This statement ensures that with overwhelming probability, a
random pure state is almost maximally entangled. The exceptional set
has measure exponentially small in a quantity essentially proportional
to the total dimension.  We will see in the next section that the
strength of this concentration gives a whole {\em large subspace} of
such states.

Whenever the reduced density matrix $\ph_A$ has near-maximal entropy,
it is also close to the maximally mixed state
$\frac{1}{d_A}\1$.  Sometimes, however, we want an even stronger estimate.
The following bound is from Appendix A of \cite{HHL03}:

\begin{lemma}[Concentration of reduced density matrices]
\label{lem:concentrationEig}
For $\ph \in_R \cP(A\ox B)$, and $0 < \e \leq 1$,
\begin{equation*}
\Pr\left\{ \lambda_{\max}(\ph_A) > \frac{1}{d_A}+\frac{\e}{d_A} \right\}
  \leq \left(\frac{10d_A}{\e}\right)^{2d_A}
  \exp\left( -d_B\frac{\e^2}{14\ln 2} \right),
\end{equation*}
and
\begin{equation*}
\Pr\left\{ \lambda_{\min}(\ph_A) < \frac{1}{d_A}-\frac{\e}{d_A} \right\}
  \leq \left(\frac{10d_A}{\e}\right)^{2d_A}
  \exp\left( -d_B\frac{\e^2}{14\ln 2} \right),
\end{equation*}
where $\lambda_{\max}$ and $\lambda_{\min}$ denote the maximal and minimal
nonzero eigenvalues of $\ph_A$, respectively.
\end{lemma}

This lemma says that the reduced state on $A$ of a random state in a
bipartite system will be close to maximally mixed in the sense that
all its eigenvalues cluster around $1/d_A$, if $d_B$ is a large enough
multiple of $d_A\log d_A / \e^2$.
In fact, when $\ph_A$ is not in the exceptional set in
Lemma \ref{lem:concentrationEig},
\bea
    (1-\e) \frac{1}{d_A}\1 \leq \ph_A \leq (1+\e) \frac{1}{d_A}\1 \, .
\label{eq:opnormbdd}
\eea
The reduced state on $B$, $\ph_B$, has the same spectrum as $\ph_A$,
and therefore $\ph_B$ will also be close to maximally mixed on its
(uniformly random) supporting subspace in a similar way.

Note that Eq.~(\ref{eq:opnormbdd}) is a statement of the concentration
of the density matrix $\ph_A$ itself, and is generally stronger than
just a bound on the von Neumann entropy as in
Theorem~\ref{thm:subsystemConvergence}.
The price paid in Lemma \ref{lem:concentrationEig} is a lesser degree
of concentration.
The main tools in proving Lemma \ref{lem:concentrationEig} also
differ from that of Theorem~\ref{thm:subsystemConvergence}.
We now state these tools, and we will use them later in the paper.
%
There are two essential ingredients.
The first is the following concentration bound, which is a slight
strengthening of Lemma II.3 from
\cite{HLSW03}, which is in turn based on Cram\'{e}r's Theorem (see
e.g.~\cite{DZ93}):
\begin{lemma}[Concentration of projector overlaps] \label{lem:concentrationTr}
For $S \in_R \cG_s(A)$, $P$ the projector onto $S$, $Q$
a fixed projector of rank $q$ in $A$,
and $0\leq \e \leq 1$,
\begin{eqnarray}
\Pr\left\{ \Tr P Q > (1+\e)\frac{qs}{d_A} \right\}
          &\leq& \exp\left( -qs\frac{\e-\ln(1+\e)}{\ln 2} \right)
           \leq  \exp\left( -qs\frac{\e^2}{6\ln 2} \right), \label{eq:proj1} \\
\Pr\left\{ \Tr P Q < (1-\e)\frac{qs}{d_A} \right\}
          &\leq& \exp\left( -qs\frac{-\e-\ln(1-\e)}{\ln 2} \right)
           \leq  \exp\left( -qs\frac{\e^2}{6\ln 2} \right). \label{eq:proj2}
\end{eqnarray}
\end{lemma}
\begin{proof}
The case $s=1$ is, in fact, a special case of \cite{HLSW03}'s Lemma II.3.
To extend to $s > 1$, let $\ket{\ph_{AB}} \in_R \cP(A\ox B)$, where
$\dim B = s$. Writing $\ph_A = \sum_i \l_i \proj{e_i}$ in its eigenbasis,
averages over $\ph_A$ can be replaced by averaging over
the independent random variables $\{ \l_i \}$ and $\{ \ket{e_i} \}$.
We can then use the convexity of the exponential function to develop
an inequality of moment generating functions. If $t\geq 0$, then
\begin{eqnarray}
\EE_{\ph_{AB}} \exp [ t s \Tr \ph_{AB} (Q \ox \1_B) ]
&=& \EE_{\ph_A} \exp [ t s \Tr \ph_A Q ] \label{eq:moment1} \\
&=& \EE_{ \{ \ket{e_j} \} } \EE_{ \{ \l_j \} }
    \exp\left[ t s \Tr \left( \sum_i \l_i \proj{e_i} Q \right) \right] \nonumber
\\
&\geq& \EE_{ \{ \ket{e_j} \} }
       \exp\left[ t s \Tr \left(
               \sum_i \EE_{ \{ \l_j \} } \l_i \proj{e_i} Q \right) \right]
\nonumber \\
&=& \EE_{ \{ \ket{e_j} \} } \exp\left[ t \Tr \left( \sum_i \proj{e_i} Q \right)
\right] \nonumber \\
&=& \EE_{S} \exp [ t \Tr P Q ]. \label{eq:moment2}
\end{eqnarray}
Here we have used that $\EE_{ \{ \a_j \} } \l_i = 1/s$, which follows from
the permutation invariance of the eigenvalue distribution.
Recall next from the proof of Lemma II.3 in \cite{HLSW03} that the
inequalities (\ref{eq:proj1}) and (\ref{eq:proj2}) for $s=1$ themselves come
from exploiting the moment generating function, in particular,
applying the general upper bound
\begin{equation}
\Pr\left\{ R > a \right\} \leq \EE_R \exp( t R )\exp(-t a)
\end{equation}
for a random variable $R$ and $t \geq 0$.
Since the left hand side of Eq.~(\ref{eq:moment1}) is the moment
generating function when $s=1$ for the larger system $AB$ with a projector
of rank $qs$, up to normalization, and Eq.~(\ref{eq:moment2}) the
moment generating function for $\Tr PQ$, the inequality
reduces the proof for $s>1$ to the $s=1$ case.
\end{proof}
%
The second tool is the existence of ``small'' fine nets in state
space, Lemma II.4 of \cite{HLSW03}.
\begin{lemma}[Existence of small nets] \label{lem:net}
For $0<\e<1$ and $\dim{\cH}=d$ there exists a set $\cN$ of
pure states in $\cH$ with
$|\cN|\leq (5/\e)^{2d}$, such that
for every pure state $\ket{\ph}\in\cH$ there exists
$\ket{\tilde{\ph}}\in\cN$ with
$\big\| \ket{\ph} - \ket{\tilde{\ph}} \big\|_2 \leq \e/2$ and
$\| \ph-\tilde{\ph} \|_1\leq\epsilon$.
(We call such a set an \emph{$\epsilon$-net}.)
\qed
\end{lemma}
%
The following is a useful generalization of Lemma
\ref{lem:net} to bipartite pure states with bounded Schmidt rank.
\begin{lemma} \label{lem:netk}
For $0<\e<1$, the set of pure states of Schmidt rank $k$ in $A \ox B$
(with dimensions $d_A$ and $d_B$) has an $\e$-net $\cN$ of size
$|\cN| \leq (10/\e)^{2k (d_A + d_B)}$.
\end{lemma}
\begin{proof}
For any Schmidt rank $k$ state $\ket{\ph}$,
there exists $U \in \UU(d)$ such that
$\ket{\ph_u} := (\1 \ox U) \ket{\ph} \in A \ox \CC^k$.
Consider an $\e/4$-net for the Hilbert space norm
$\| \cdot \|_{2}$ on $A \ox \CC^k$, and
let $\ket{\tilde{\ph}_u}$ be a net point for $\ket{\ph_u}$, with Schmidt
decomposition
$\ket{\tilde{\ph}_u} = \sum_i \sqrt{\lambda_i} \ket{e_i} \ox \ket{f_i}$.
Then, $(\1 \ox U^\dag) \ket{\tilde{\ph}_u}
  = \sum_i \sqrt{\lambda_i} \ket{e_i} \ox (U^\dag \ket{f_i})$
is within a distance $\e/4$ of $\ket{\ph}$ in $\|\cdot \|_2$.
Now, take an $\e/4$-net for $\| \cdot \|_2$ on $B$ and let
$\ket{\tilde{f}_i}$ be the net point for $U^\dag \ket{f_i}$.
It is straightforward to check that $\ket{\tilde{\tilde{\ph}}} := \sum_i
\sqrt{\lambda_i} \ket{e_i} \ox \ket{\tilde{f}_i}$ is within $\e/4$ of
$(\1 \ox U^\dag) \ket{\tilde{\ph}}$ in $\|\cdot\|_2$. By the triangle
inequality, $\ket{\tilde{\tilde{\ph}}}$ is within $\e/2$ of
$\ket{\ph}$ in $\| \cdot \|_2$, and within $\e$ in the trace norm
$\| \cdot \|_1$.
Altogether, the total number of net points is
$(10/\e)^{2 k d_A} (10/\e)^{(2 d_B) k}$, as claimed.
\end{proof}

\vspace{1ex}

We end the section with another Lipschitz constant estimate that will
be useful when studying the separability of random states.
\begin{lemma} \label{lem:Lip2}
  Let $\ket{\ph} \in A\ox B$. Then
  the Lipschitz constant of the function $f(\ph) = \sqrt{\Tr\ph_A^2}$ is upper
  bounded by $2$.
\end{lemma}

\begin{proof}
  Choose a basis $\{\ket{e_i}\ket{f_j} \}$ of $A \ox B$,
  and let $\ket{\ph} = \sum_{ij} \ph_{ij}\ket{e_i}\ket{f_j}$.
  Since $\Tr \ph_A^2$ is nonincreasing under dephasing,
  \begin{equation*}
    f(\ph)^2 = \Tr\r^2 \geq \sum_i \bra{e_i} \ph_A \ket{e_i}^2 =
    \sum_i \Bigl( \sum_j |\ph_{ij}|^2 \Bigr)^2 =: \tilde{f}(\ph)^2
  \end{equation*}
  with equality if $\{\ket{e_i}\}$ are the eigenvectors of $\ph_A$.
  The Lipschitz constant of $\tilde{f}(\ph)$ is easily seen to
  be bounded by $2$: simply calculate the length of the gradient
  and use standard inequalities.
  We now apply the trick that proves Lemma \ref{lem:E-Lipschitz}.
  For any two pure states $\ph$ and $\psi$ with $f(\ph) \geq f(\psi)$,
  choose $\{\ket{e_i}\}$ to be the eigenbasis of $\ph$.  Then,
  $\tilde{f}(\ph)=f(\ph)$,
  $\tilde{f}(\psi) \leq f(\psi)$, and
  \begin{equation*}
    f(\ph) - f(\psi)
        \leq  \tilde{f}(\ph) - \tilde{f}(\psi)
        \leq 2\bigl\| \,\ket{\ph}-\ket{\psi}\, \bigr\|_2.
  \end{equation*}
\end{proof}

The following sections will demonstrate the power of the above basic
concentration statements and net construction for the understanding of
generic entanglement.

\section{Maximally entangled subspaces} \label{sec:subspace}

In this section, we put together the insights from the previous
section to show, that a
large subspace of appropriate dimension, chosen at random, will with
high probability contain only near-maximally entangled states. The
relationship between concentration of measure and statements about
large subspaces play an important role in \cite{MS86}. The reader is
also encouraged to compare our result with Theorem 3.19 of \cite{L01}.

\begin{theorem}[Entangled subspaces] \label{thm:subspace}
Let $A$ and $B$ be quantum systems of dimension $d_A$ and $d_B$,
respectively, for $d_B \geq d_A \geq 3$.  Let $0<\a<\log d_A$.  Then
there exists a subspace $S \subset A \ox B$ of dimension
\begin{equation} \label{eqn:sizeSubspace}
s = \left\lfloor d_A d_B \frac{\G\a^{2.5}}{(\log d_A)^{2.5}} \right\rfloor
\end{equation}
such that all states $\ket{\ph} \in S$ have entanglement at least
\begin{equation}
E(\ph) = S(\ph_A) \geq \log d_A - \a - \b,
\label{eqn:entanglementbound}
\end{equation}
where $\b = \frac{1}{\ln 2}\frac{d_A}{d_B}$ is as in Lemma
\ref{lem:entropySubsystem} and $\G$ is an absolute constant which may
be chosen to be $1/1753$.
In fact, the probability that a random subspace of dimension $s$
will not have this property is bounded above by
\begin{equation}
\left(\frac{15 \log d_A}{\a}\right)^{\! \! 2s}
  \exp\left( - \frac{(d_A \, d_B \, - 1) \, \a^2}
                    {32 \pi^2 \ln 2 \, (\log d_A)^2} \right).
\label{eq:subspaceexcept}
\end{equation}
\end{theorem}

\begin{proof}
Let $S$ be a random subspace of $A\ox B$ of dimension $s$.
Let $\cN_S$ be an $\e$-net for states on $S$, for
$\e=\a/(\sqrt{8}\log d_A)$. In fact, since we may think of $S$ as $U S_0$,
with a \emph{fixed} subspace $S_0$ and a Haar-distributed unitary $U$,
we can fix the net $\cN_{S_0}$ on $S_0$ and let $\cN_S=U \cN_{S_0}$,
where $\cN_{S_0}$ is chosen using Lemma \ref{lem:net}.
Given $\ket{\ph} \in S$, we can choose
$\ket{\tilde{\ph}} \in \cN_S$ such that
$\big\| \ket{\ph} - \ket{\tilde\ph} \big\|_2 \leq \e/2$.
By the Lipschitz estimate, Lemma \ref{lem:E-Lipschitz}, this implies
that $|S(\ph_A) - S(\tilde{\ph}_A)| \leq \a/2$.
We can then estimate
\begin{eqnarray}
\Pr\Big\{ \inf_{\ket{\ph} \in S} S(\ph_A)
  < \log d_A - \a - \b \Big\} \label{eqn:subspaceProb}
&\leq& \Pr\Big\{  \min_{\ket{\tilde{\ph}}\in\cN_S}
  S(\tilde{\ph}_A) < \log d_A - \a/2 - \b \Big\} \\[0.5ex]
&\leq& |\cN_S| \; \Pr\big\{ S(\ph_A) < \log d_A - \a/2 - \b \big\} \nonumber
\\[0.5ex]
&\leq& \left(\frac{15 \log d_A}{\a}\right)^{\! \! 2s}
  \exp\left( - \frac{(d_A \, d_B \, - 1) \, \a^2}{32 \pi^2 \ln 2 \, (\log
d_A)^2} \right). \nonumber
\end{eqnarray}
This proves the upper bound on the probability that the randomly
selected subspace $S$ will not satisfy the large entanglement
requirement.
If this is smaller than $1$, a subspace with the stated properties
exists; this can be secured by requiring
\begin{equation}
s < \frac{(d_A d_B - 1)\a^2}
         {438(\log d_A)^2 \log(15 \log d_A/\a)} \,.
\label{eq:dumbound}
\end{equation}
A less tight but simpler expression can be obtained.
By using $\log x \leq \sqrt{\frac{16}{15}x}$ for $x\geq 15$, we get
$\log(15\log d_A/\a) \leq 4 \sqrt{\frac{\log d_A}{\a}}$,
because $\a < \log d_A$.
Now, if we are to replace the denominator in Eq.~(\ref{eq:dumbound})
by this new expression and still aim to make a non-vacuous statement
(i.e. that $s\geq 2$), then, keeping in mind that $\a < \log d_A$,
we find that $d_A d_B \geq 3505$, so that
$d_A d_B - 1 \geq \smfrac{3504}{3505} d_A d_B$, leading to
Eq.~(\ref{eqn:sizeSubspace}).
\end{proof}


\medskip

If $\a+\b$ is small, we can obtain another useful characterization of
all the states in $S$: that they are all close to maximally entangled
states.

\begin{corollary} \label{cor:subspace-fidelity}
Every pure state $\ket{\ph}\in S$ constructed in Theorem
\ref{thm:subspace} is close to a maximally entangled state $\ket{\Phi}
\in A \ox B$:
\begin{equation*}
  F(\ph,\Phi)      \geq 1-\sqrt{2(\a+\b)} \, , \qquad
  \| \ph-\Phi \|_1 \leq \sqrt[4]{16(\a+\b)} \, .
\end{equation*}
\end{corollary}
\begin{proof}
The relative entropy between $\ph_A$ and the maximally mixed state is
given by
\begin{equation*}
D\left(\ph_A \left\|\smfrac{\1}{d_A} \right.\right)
                  = \log d_A - E(\ph) \leq \a+\b \,.
\end{equation*}
Hence, by Pinsker's inequality (see \cite{OhyaP93,SW01}),
$\| \ph_A - \1/d_A \|_1 \leq \sqrt{2(\a+\b)}$.
Using a well-known relation between trace distance and fidelity
\cite{Fuchsv99}, we obtain
$F( \ph_A, \1/d_A) \geq 1-\sqrt{2(\a+\b)}$.
By Uhlmann's Theorem~\cite{Uhlmann76,Jozsa94}
this means that $\ph$ is indeed close
to a purification of the maximally mixed state, i.e., a maximally
entangled state: there exists a maximally entangled state $\Phi$
such that $F(\ph,\Phi) \geq 1-\sqrt{2(\a+\b)}$, and hence,
invoking~\cite{Fuchsv99} once more,
$\| \ph - \Phi \|_1 \leq \sqrt[4]{16(\a+\b)}$.
\end{proof}

\section{Correlation measures for random states} \label{sec:measures}

In this section, we consider correlation properties of rank-$s$ random
states with distributions induced by partial tracing (see Definition
\ref{def:randommixed}).
Our study was motivated by some surprising properties of the maximally
mixed states on the random subspaces discussed
in the previous section.
Since the spectrum of a rank-$s$ random state is likely to be
almost flat, the two types of mixed states are very similar
asymptotically, at least for the purposes of our investigation.
Thus, after a full discussion on the asymptotic correlation properties
of rank-$s$ random states, we derive, as corollaries, asymptotic
correlation properties of maximally mixed states on random
subspaces.

\subsection{Some measures of correlation for quantum states}

Consider interconversions between copies of some state $\s_{AB}$ and EPR pairs
by local operations and (two-way) classical
communications (LOCC) in the limit of many copies.
The number of EPR pairs needed per copy of $\s$ created is defined
to be the entanglement cost~\cite{HHT00},
$E_c(\s_{AB}) = \lim_{n\rar\infty} \frac{1}{n} E_f(\s_{AB}^{\ox n})$,
where
\begin{equation}
E_f(\s_{AB})
= \min_{\sum_i p_i \proj{\ph^{i}}
= \s_{AB}} \sum_i p_i S(\ph_A^i)
\label{eq:ef}
\end{equation}
is the entanglement of formation~\cite{BDSW96}.  $\s_{AB}$ is said to
be {\em separable} if $E_f(\s_{AB}) = 0$.
It is proved in \cite{GB03} that any $d$-dimensional state $\s$ is
separable if
\bea
    \Tr \s^2 \leq 1/(d-1) \,.
\label{eq:separable}
\eea
The number of EPR pairs that can be extracted per copy of $\s_{AB}$ is
given by the entanglement of distillation, $E_d$~\cite{BDSW96}.
One can also quantify the amount of secret key $K(\s)$ distillable
against an eavesdropper holding the purification of the state
(see~\cite{D03,DW03b,DW03c} and references therein), and the distillable
``common randomness'' ${\rm CR}(\s)$~\cite{AC98,DW03a} (discounting at the end
of the protocol the amount of communication used).
When the communication is restricted to one direction, say, from
$A$ to $B$, we can define the corresponding distillable correlations
$E_d^\rar(\s)$, $K^\rar(\s)$, and ${\rm CR}^\rar(\s)$.
In particular, it is proved in \cite{DW03a} that
\begin{equation} \label{eq:crinfo}
{\rm CR}^\rar(\s) = \lim_{n\rar\infty} \frac{1}{n} I^\rar(\s^{\ox n}),
\end{equation}
where $I^\rar(\s)$ is the {\em maximum Holevo
quantity}~\cite{H73} of the reduced ensemble of states in $B$
induced by a local measurement in $A$, if the initial state is $\s$.
One can also formally define quantum mutual information
\begin{eqnarray} \label{eq:sab}
S(A:B)_\s = S(\s_A) + S(\s_B) - S(\s_{AB}) \,,
\end{eqnarray}
by analogy to Shannon's classical quantity.
The various measures are related by many known inequalities:
\bea
    E_d(\s) & \leq & E_c(\s) \leq E_f(\s) \,,
\label{eq:irrev-add}
\\
    E_d(\s) & \leq & \smfrac{1}{2} \, S(A:B)_\s \,,
\label{eq:squash}
\\
    E_d(\s) & \leq & K(\s) \leq {\rm CR}(\s) \leq S(A:B)_\s \,,
\label{eq:trivial2way}
\\
    E_d^\rar(\s) & \leq & K^\rar(\s) \leq {\rm CR}^\rar(\s) \,,
\label{eq:trivial1way}
\eea
as well as the trivial bounds $E_d^\rar(\s) \leq E_d(\s)$ etc.
Most of these inequalities follow directly from the operational
definitions.
Equation (\ref{eq:squash}) was proved in \cite{CW03}, exploiting the fact
that the right hand side is an upper bound on the ``squashed entanglement.''
The rightmost inequality in Eq.~(\ref{eq:trivial2way}) can easily be
proved by generalizing the classical case~\cite{AC98}.

As alluded to earlier, our investigation of correlation for random states was
motivated by thinking about states on a maximally
entangled subspace $S$ produced by Theorem \ref{thm:subspace}.
Any state on $S$, pure or mixed, has entanglement of
formation at least $\log d_A - \a -\b$.  Meanwhile, for the
maximally mixed state $\r_{AB} = \int_S \proj{\ph} \; d\ph$ on $S$,
$S(\r_{AB}) = \log s$ can be very high.
Taking $\a \leq 1$ and $s$ equal to the value given by
\eqn{sizeSubspace} leads to a strong upper bound on the mutual information:
\begin{eqnarray}
S(A:B)_\r \leq 2.5 \log\log d_A - \log(\G\a^{2.5}) +1 \,.
\label{eqn:mutualBound}
\end{eqnarray}
It follows from Eqs.~(\ref{eq:squash})--(\ref{eq:trivial1way}) that
all of $E_d(\r)$, $K(\r), {\rm CR}(\r)$ are small.
In particular, $E_f(\r) \gg E_d(\r)$, so that either $E_f(\r) \gg
E_c(\r)$ or $E_c(\r) \gg E_d(\r)$.  In the first case, $\r_{AB}$ is a
(rather drastic) counter-example to the additivity conjecture for
the entanglement of formation:
$\forall \s \; E_f(\s_{AB}^{\ox n}) = n E_f(\s_{AB})$ (see, for
example,~\cite{S04}).  In the second case, the preparation of $\r_{AB}$
is near-maximally irreversible, making it a kind of entanglement black hole;
preparing $\r_{AB}$ requires nearly as much entanglement as the most
highly entangled state even though no useful entanglement can be extracted
from it.
Moreover, for most values of $s$, the gap $E_f(\r) \gg
E_d(\r)$ is generic.


Each of these states $\r_{AB}$ also provides an example of a quantum
state that is more ``entangled'' than it is ``correlated'' --- a
hitherto unseen effect. At the very least, this reveals that ``dividing''
the correlations of a quantum state into
entanglement and classical parts is problematic,
since here we find a measure of
entanglement that can exceed the combined quantum and classical
correlations. It is interesting to note, however, that if one replaces
the entanglement of formation by operational measures related to the
entanglement of distillation, this decomposition of correlation into
quantum and classical parts becomes possible, as demonstrated in
\cite{OHHHH03}.

\subsection{Analysis of correlation measures for rank-$s$ random states}

Throughout this section we select random states according to the
prescription $\r \in_R \cS_s(A \ox B)$, with $d_B \geq d_A \geq 3$.
Since the statements of the theorems are rather technical, we will
begin by sketching a rough outline of the results to come.
We will frequently need to make statements
conditioned on the additivity
conjecture for $E_f$ or, equivalently, $I^\rar$ (see~\cite{KW04}
for the equivalence).
To simplify the discussion, from now on
we will indicate that a statement is true
conditioned on the conjecture by marking it with the symbol $*$,
either as a superscript at the end of a sentence or above a mathematical
symbol: $\stackrel{*}{=}$, for example.
As a start, we confirm the gap between $E_f(\r)$ and $S(A:B)_\r$ that
was discussed earlier for random maximally mixed states.
Our findings are summarized in the Table \ref{table:high-rank}.

\bigskip

%
%
\begin{table}[h]
\begin{center}
\begin{tabular}{|c|l|c|}\hline
\multicolumn{3}{|c|}{\textbf{Properties of high-rank random states:}} \\
\multicolumn{3}{|c|}{
  $\e > 0, \, d_A = d, \, d_B = d \log d, \, s = d^2/(\log d)^6$
} \\[0.3ex] \hline
Correlation Measure & Value & Likelihood \vholder{2.4ex} \\[0.3ex] \hline\hline
$E_f \; (\stackrel{*}{=} E_c)$
  & $\geq \log d - \e$          
  & $d^2$-likely    \vholder{2.4ex} \\[0.3ex] \hline
$S(A:B) \; (\geq {\rm CR} \geq K, \, \geq 2 E_d)$
  & $\leq 7 \log \log d$        
  & $d^4$-likely    \vholder{2.4ex} \\[0.3ex] \hline
$I^\rar \; (\stackrel{*}{=} {\rm CR}^\rar \geq K^\rar \geq E_d^\rar)$
  & $\leq \e$
  & $d^2$-likely    \vholder{2.4ex} \\[0.3ex] \hline
\end{tabular}
\end{center}
\caption{Properties of high-rank random states. For the specific choice
of parameters made here, the gap between the entanglement of formation
and the measures of distillable correlation is basically as large
as is consistent with the entropy scale of the system.}
\label{table:high-rank}
\end{table}

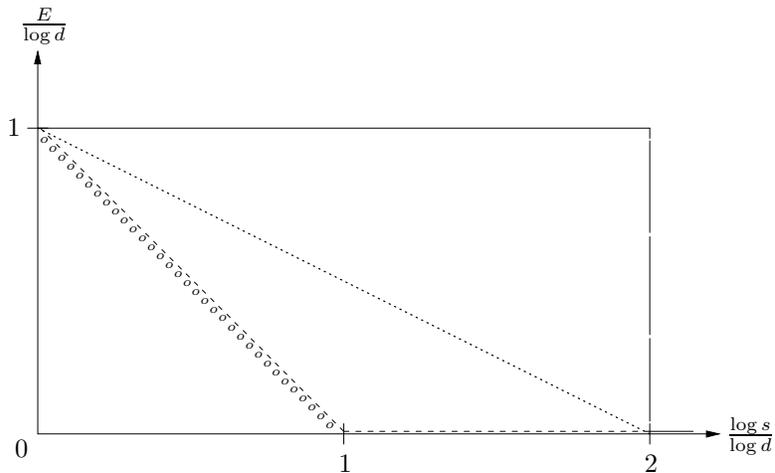
\begin{figure}[h]

\setlength{\unitlength}{0.00053333in}
\begingroup\makeatletter\ifx\SetFigFont\undefined%
\gdef\SetFigFont#1#2#3#4#5{%
  \reset@font\fontsize{#1}{#2pt}%
  \fontfamily{#3}\fontseries{#4}\fontshape{#5}%
  \selectfont}%
\fi\endgroup%
{\renewcommand{\dashlinestretch}{30}
\begin{picture}(7324,4425)(0,-10)
\thicklines
\blacken\thinlines
\path(405.000,4005.000)(375.000,4125.000)(345.000,4005.000)(405.000,4005.000)
\path(375,4125)(375,375)(7050,375)
\blacken\path(6930.000,345.000)(7050.000,375.000)(6930.000,405.000)(6930.000,345.000)
\path(275,3375)(475,3375)
\path(375,3375)(6375,3375)
\path(6375,3375)(6375,3275)
\path(6375,3250)(6375,2350)
\path(6375,2310)(6375,1350)
\path(6375,1310)(6375,500)
\path(6375,400)(6800,400)
%
\dottedline{60.000}(375,3375)(6375,375)
\path(3375,475)(3375,275)
\path(6375,475)(6375,275)
\dottedline[${}_{{}^o}$]{125}(450,3230)(3250,430)
\dashline{60}(400,3375)(3375,400)
\dashline{60}(3375,400)(6375,400)
\put(75,3300){$1$}
\put(3325,0){$1$}
\put(6325,0){$2$}
\put(7125,300){$\frac{\log s}{\log d}$}
\put(200,4325){$\frac{E}{\log d}$}
\put(150,150){$0$}
\end{picture}
}

\caption{Illustration of the asymptotic ($d\rar\infty$) behavior
  of entanglement $E$ versus rank $s$ of random states in $\CC^d\ox\CC^d$,
  with all quantities normalized over $\log d$, the entropy scale of the
  system. The solid line is the entanglement of formation, dropping
  sharply from $1$ to $0$ at the threshold $\frac{\log s}{\log d} \sim 2$.
  The dotted line is the upper bound on distillable entanglement
  from Theorem~\ref{thm:random-EoF} and Eq.~(\ref{eq:squash}),
  and the circled line is a lower bound
  on the one-way distillability via the hashing inequality:
  $E^\rar_d(\rho_{AB}) \geq S(\rho_B)-S(\rho_{AB})$~\cite{DW03b,DW03c}.
  Finally, the dashed line is the one-way distillable
  common randomness$^*$ from
  Theorem~\ref{thm:oneway-bounds}. Hence,
  the dashed line also represents the one-way distillable entanglement.$^*$
}
\label{fig:asymptotics}
\end{figure}

In fact, we are able to determine much more. Other
than when $s$ is almost exactly equal to $d_A d_B$, we can compute
excellent approximations to both $E_f$ and $I^\rar$.
Assuming the additivity conjecture, that is sufficient to
calculate the entanglement cost and one-way distillable entanglement of
rank-$s$ random quantum states.
Figure \ref{fig:asymptotics} illustrates the situation when
$d_B = d_A = d$ becomes large; it plots
the normalized entanglement of formation, squashed entanglement
and coherent information (see \cite{DW03b,DW03c} for the relevant lower bound)
against the normalized entropy, which is essentially
the rank, of the likely random states from $\cS_s(A\ox B)$.

\medskip

\pagebreak
\begin{theorem}
\label{thm:random-EoF}
Let $\rho \in_R \cS_s(A\ox B)$, with $d_B \geq d_A \geq 3$,
and $0 < \a < \log d_A$. Then:
\begin{enumerate}
\item If $s< d_A d_B (\log d_A)^{-2.5} \; \G\a^{2.5}$,
        then it is $d_A d_B$-likely that
    $E_f(\r_{AB}) \geq \log d_A - \a -\b$.  \\[1ex] The parameters
    $\a,\b,\G$ are the same as in Theorem \ref{thm:subspace}.
\item If $s > d_A d_B (\log d_A)^2 (6 \log d_B - 4 \log \e)
       \frac{14 \ln 2}{\e^2}$, then it is $s$-likely
       that $E_f(\rho) \leq \e$.
\item If $s > 6(d_A d_B)^2$,
    then it is $\smfrac{s}{(d_A d_B)^2}$-likely that $\rho$ is separable.
\item If $s < d_A d_B$, it is $s d_A d_B$-likely that
    $S(A:B)_\r \leq \log d_A + \log d_B - \log s + \a + \b_1$
    for $\b_1 = \smfrac{1}{\ln 2}\smfrac{s}{d_A d_B}$. \\
      If $s > d_A d_B$, it is $s d_A d_B$-likely that
    $S(A:B)_\r \leq \a + \b_2$ for
        $\b_2 = \smfrac{1}{\ln 2}\smfrac{d_A d_B}{s}$.
\item
$E_d^{\rar}(\r)$, $K^{\rar}(\r)$, ${\rm CR}^{\rar}(\r)$
$E_d(\r)$, $K(\r)$, ${\rm CR}(\r)$
share the same upper bound as $S(A:B)_\r$, due to
Eqs.~(\ref{eq:squash})--(\ref{eq:trivial1way}).
\end{enumerate}
Bounds on the probabilities $P_E$ of the various exceptional sets
are given in the proof.
\end{theorem}
\begin{proof}
\begin{enumerate}
\item
By the uniqueness of the unitarily invariant measure on the
Grassmannian, the support of $\r$ {\em is} a random $s$-dimensional
subspace $S$.  Since $s$ satisfies the condition of \eqn{sizeSubspace}
in Theorem \ref{thm:subspace}, the claim follows from
Eq.~(\ref{eq:ef}), with $P_E$ given in \eq{subspaceexcept}.
\item
We apply Lemma \ref{lem:concentrationEig} to $(A\ox B) \ox
\CC^{s}$ and choose $\e'$ so that $(1-\e') \smfrac{\1}{d_A d_B} \leq
\r \leq (1+\e') \smfrac{\1}{d_A d_B}$ is $s$-likely, in which case $\r =
(1-\e') \smfrac{\1}{d_A d_B} + \e' \r'$, for some state $\rho'$.  \\[1ex]
If we choose $\e' = \e/\log d_A$, then
by the convexity of $E_f$, $E_f(\r) \leq \e' E_f(\r') \leq \e$, and
$P_E \leq 2 \bigl(10 d_A d_B (\log d_A)^2 /\e^2 \bigr)^{2 d_A d_B}
\exp\bigl(-\smfrac{s}{(\log d_A)^2} \smfrac{\e^2}{14 \ln 2} \bigr)$.
\item
Setting $d=d_A d_B$,
we shall bound $\Tr \r^2$ by $\frac{1}{d}+\frac{1}{d^2}$
and use Levy's Lemma
to estimate the probability that this occurs;
then by \eq{separable} we are done.
This requires Lemma \ref{lem:Lip2} for the upper bound of $2$ on the
Lipschitz constant of $f(\rho) = \sqrt{\Tr \r^2}$, and a result
from~\cite{L78} which says
$\EE\Tr\r^2 = \frac{d+s}{ds+1}$.
Noting that $\Tr\r^2 \geq 1/d$, an argument as in the proof of
Theorem \ref{thm:subsystemConvergence} then implies that the median
$m(\Tr\r^2) \leq {1}/{d} + {2}/{s}$. Then, by the definition of the
median, $m(f) = \sqrt{m(\Tr\r^2)} \leq \sqrt{1/d+2/s} \leq
1/\sqrt{d} + \sqrt{d}/s$.
Since, on the other hand,
$\sqrt{1/d+1/d^2} \geq \sqrt{1/d}+1/(3d^{3/2})$,
and, assuming $s \geq 6d^2 $, Lemma~\ref{lem:Levy} yields
\begin{equation*}
  P_E \leq \Pr\left\{ f > \sqrt{\frac{1}{d}+\frac{1}{d^2}} \right\}
      \leq \Pr\left\{ f > m(f) + \frac{1}{6d^{3/2}} \right\}
      \leq \exp\left( -\frac{ds-1}{493d^3} \right),
\end{equation*}
which is less than one.
\item
Using $S(\r_A) \leq \log d_A$ and $S(\r_B) \leq \log d_B$, it
will be sufficient to lower bound $S(\r_{AB})$.
If $s < d_A d_B$, apply Theorem \ref{thm:subsystemConvergence} to the
bipartite system $\CC^s \ox A \ox B$, $S(\r) \geq \log s - \a - \b_1$ with
$P_E \leq \exp \bigl( -(s d_A d_B - 1) C_3 \a^2 / (\log s)^2
\bigr)$.
If $s > d_A d_B$, the same theorem gives $S(\r) \geq \log(d_A d_B) -
\a - \b_2$, this time with
$P_E \leq \exp \bigl( -(sd_A d_B -1) C_3 \a^2 / (\log d_A d_B)^2 \bigr)$.
Note that the concentration effect is achieved via a large $d_A d_B$,
and very little is required of $s$.
\end{enumerate}
\end{proof}
\begin{remark}
As a particular consequence of Theorem~\ref{thm:random-EoF}, Parts
1 and 2, note that $E_f$ drops rapidly from near-maximal to near-minimal
when the rank $s$ of the random state under consideration increases
over the threshold $s \sim d_A d_B$. This very rapid drop,
a prominent feature in Figure \ref{fig:asymptotics}, can be interpreted
as another instance of the ``locking effect'' for $E_f$ pointed
out in~\cite{HHHO04}: let Alice and Bob each hold systems of $n$
qubits each, and provide them with a random state $\rho$ of rank
$s = 2^{2n}/\poly(n)$ according to
Theorem~\ref{thm:random-EoF}, Part 1, so that $E_f(\rho) \geq n-1$
is likely. Then let Alice (and/or Bob) trace out $O(\log n)$
of their qubits such that the resulting random state $\rho'$
satisfies the criteria of Part 2 of the theorem.
Since it is likely that $E_f(\rho') \leq \e$ for arbitrarily small
$\e$, we conclude that tossing away only $O(\log n)$
qubits can change the entanglement of formation of a $2n$-qubit
state by almost $n$.
What is more, we have identified a regime where
this effect is the overwhelmingly likely rule.
\end{remark}

The upper bounds on the various correlation measures in Parts
4 and 5 of Theorem \ref{thm:random-EoF} are negligible compared to
$E_f(\rho)$ for $s \sim d_A d_B / \polylog(d_A)$,
but still growing as $O(\log\log d_A)$.
We now prove a {\em vanishing} upper bound for the quantity
$I^{\rar}(\r)$ and random states of rank larger than $d_B$.
The demonstration is closely related to the proof that random encodings can
be used for quantum data hiding~\cite{HLSW03,HLS04}.
\begin{theorem} \label{thm:oneway-bounds}
Let $\rho \in_R \cS_s(A\ox B)$ be a random state of rank $s$.
Then, for $\e \leq 1/3$ and $d_B \geq d_A$,
\begin{equation*}
\Pr\left\{ I^\rar(\rho) > 5\e \right\}
  \leq 2 \left( \frac{20 d_B^2}{\e} \right)^{4d_B}
                           \exp\left( -s\e^2/17  \right).
\end{equation*}
Thus it is likely that $I^{\rar}(\r)$ is asymptotically vanishing as
long as the rank $s$ of $\rho$ is sufficiently bigger than $d_B
\log d_B$.
%
\eqs{crinfo}{trivial1way} therefore imply the same
vanishing upper bound on $E_d^{\rar}(\r)$, $K^{\rar}(\r)$
and ${\rm CR}^{\rar}(\r)$.$^*$

\noindent On the other hand, for $s/\e \leq d_B \leq \e s d_A$, we have
\begin{equation*}
\Pr\left\{ \bigl| I^{\rar}(\r) - (\log d_B-\log s) \bigr| > 2\e \right\}
    \leq 2 \left( \frac{15 \log s}{\e} \right)^{2 d_A}
              \exp\left( -\frac{(sd_B-1)\e^2}{32 \pi^2 \ln 2(\log s)^2}\right).
\end{equation*}
In other words, as long as the above constraints and
$d_A < s d_B \frac{\G\e^{2.5}}{(\log s)^{2.5}}$ hold,
it is $sd_B$-likely that $I^{\rar}(\r) = \log d_B - \log s \pm 2\e$.
\end{theorem}

\begin{proof}
We start with the explicit expression for $I^\rar(\rho)$ proved
in~\cite{DW03a}:
\begin{equation}
\label{eq:I-rar}
I^\rar(\rho) := S(\rho_B) - \min_{M\text{ POVM on }A}
  \sum_i \Tr(\rho_A M_i) \;
  S \biggl( \frac
    {\Tr_A \bigl(\rho\,(M_i \hspace{-0.3ex} \ox \hspace{-0.3ex} \1)\bigr)}
    {\Tr(\rho_A M_i)} \biggr)
\end{equation}
where the right hand side is the Holevo quantity~\cite{H73} on Bob's reduced
ensemble of states labeled by the measurement outcome. Without loss of
generality,
all $M_i$ are of rank one.
For the first part of the theorem, we will show
that it is likely that for every rank-one projector
$P$ acting on $A$, the corresponding projected state on $B$,
\bea
    \sigma = \frac{\Tr_A\bigl( \rho(P\ox\1) \bigr)}
        {\Tr\bigl( \r \, (P\ox\1) \bigr)}
\label{eq:sig}
\eea
is close to maximally mixed, so that for every POVM on $A$, the
difference in Eq.~(\ref{eq:I-rar}) is small.
\par
Since $\rho=\Tr_{\CC^s} \ph$ for $\ph\in_R A\ox B\ox \CC^s$,
we have, for rank-one projectors $P$ and $Q$,
$\Tr\bigl( \rho(P\ox Q) \bigr) = \Tr\bigl( \ph(P\ox Q\ox \1) \bigr)$,
so Lemma~\ref{lem:concentrationTr} gives us
\begin{equation}
\Pr\left\{
     \left| \Tr\bigl(\r \, (P\ox Q)\bigr)-\frac{1}{d_Ad_B} \right|
                                      > \frac{\e/2}{d_A d_B} \right\}
             \leq 2\exp\left( -s\e^2/17 \right) .
     \label{eq:expBoundI}
\end{equation}
Now, Lemma~\ref{lem:net} gives us $\frac{\e}{4 \, d_A d_B}$-nets for the pure
states (rank one projectors) on $A$ and $B$, of cardinality
$({20 \, d_A d_B}/{\e})^{\!2d_A}$ and
$({20 \, d_A d_B}/{\e})^{\!2d_B}$,
respectively. Hence, by the union bound and triangle inequality,
\begin{equation*}
  \Pr\left\{ \exists P,Q \
       \left| \Tr\bigl(\r \, (P\ox Q)\bigr)-\frac{1}{d_Ad_B} \right|
                                          > \frac{\e}{d_Ad_B} \right\}
  \leq 2 \left(\frac{20 d_A d_B}{\e}\right)^{2(d_A+d_B)}
     \exp\left(-s\e^2/17 \right).
\end{equation*}
If this event does not occur, then for every rank-one projector $P$,
\begin{equation*}
  (1-\e)\frac{1}{d_B}\1 \leq \Tr_A \bigl( \r \, (P\ox\1) \bigr) \leq
    (1+\e)\frac{1}{d_B}\1 \,,
\end{equation*}
and the post-measurement state $\s$ as defined in \eq{sig} satisfies
\begin{equation*}
  (1-3\e)\frac{1}{d_B}\1 \leq \sigma \leq (1+3\e)\frac{1}{d_B}\1,
\end{equation*}
which in turn easily implies
\begin{equation*}
  S(\sigma) \geq \log d_B - \frac{3\e}{\ln 2} \geq \log d_B - 5\e,
\end{equation*}
by the operator monotonicity of $\log$.
Putting this, using $d_B \geq d_A$
and substituting $S(\rho_B) \leq \log d_B$ into \eq{I-rar} completes
the argument.

For the second statement, we will use an alternative argument
based on the entangled subspaces of Theorem~\ref{thm:subspace}.
To begin with, there exists a $\ph\in_R {\cal P}(A\ox B\ox \CC^s)$
such that $\rho = \Tr_{\CC^s}\ph$.
Hence, Theorem~\ref{thm:subsystemConvergence} informs us that
\begin{equation}
\Pr\left\{ S(\rho_B) < \log d_B - \e - \frac{1}{\ln 2}\frac{d_B}{s d_A} \right\}
  \leq \exp\left( -\frac{(sd_Ad_B-1)C_3\e^2}{(\log d_B)^2} \right).
  \label{eq:theentropy}
\end{equation}
On the other hand, consider the post-measurement state $\sigma$
on $B$ as in \eq{sig} --- it can clearly be written as the corresponding
post-measurement (pure) state $\psi$ on $B\ox\CC^s$, reduced to $B$:
$\sigma = \psi_B$, and $S(\sigma) = E(\psi)$. But $\ket{\psi}$
lies in the supporting subspace of $\Tr_A\ph$, which is a random
subspace of dimension $d_A$ in $B\ox\CC^s$.
Hence we can apply Theorem~\ref{thm:subspace}, telling us
\begin{equation*}
\Pr\left\{ \exists \text{post-meas. state }\sigma,\ S(\sigma)
               < \log s - \e -\frac{1}{\ln 2}\frac{s}{d_B} \right\}
    \leq \left( \frac{15 \log s}{\e} \right)^{2 d_A}
              \exp\left( -\frac{(sd_B-1)\e^2}{32\pi^2\ln 2(\log s)^2}\right).
\end{equation*}
Since this dominates the bound in \eq{theentropy},
we will be done if we just insert
our entropy bounds,
$\log d_B -\e - \smfrac{1}{\ln 2}\smfrac{d_B}{s d_A}
 \leq S(\rho_B) \leq \log d_B$ and
$\log s -\e - \smfrac{1}{\ln 2}\smfrac{s}{d_B}
 \leq S(\sigma) \leq \log s$,
into \eq{I-rar} and respect the dimension
constraints we inherit.
\end{proof}


\bigskip
We finish this subsection by considering a more qualitative aspect of
entanglement of a state $\rho$ on $A\ox B$, \emph{one-copy
(pseudo-)distillability}, meaning that there exist two-dimensional
projectors $P$ and $Q$ on $A$ and $B$ respectively such that
\bea
    \sigma = \frac{(P \ox Q) \rho (P \ox Q)}{\Tr(P \ox Q) \rho}
\label{eq:sig2}
\eea
has {\em partial transpose} that is not positive
semidefinite~\cite{P96} (NPT). The motivation is that in this case
$\sigma$ is effectively a two-qubit state, and $\s$ is distillable if
it is NPT, and separable if it is not (PPT)~\cite{HHH97}.  Furthermore,
$\rho$ is distillable if and only if $\rho^{\ox n}$ is one-copy
distillable for some $n$~\cite{HHH98}.

\begin{theorem} \label{thm:one-copy-undistillable}
Let $\rho \in_R \cS_s(A\ox B)$ be a random state of rank $s$
with $d_B \geq d_A$.
Then,
\begin{equation*}
\Pr\left\{ \rho \text{ one-copy distillable} \right\}
  \leq 2 (10d_B)^{16d_B} \exp\left( -s/600\ln 2 \right)
\end{equation*}
In particular, once $s > 7000 d_B\log(10d_B)$,
$\rho$ is likely to be one-copy undistillable.
\end{theorem}

\begin{proof}
We will show that $\forall P \ox Q$, $\sigma$ in \eq{sig2} is likely
to be separable, using a characterization from~\cite{GB03} that
$\sigma$ is separable if $\| \sigma-\frac{1}{4}\1 \|_\infty \leq
\frac{1}{8}$.

To show the above, fix any Schmidt-rank two state $\ket{\ps}$ on $A \ox B$.
Since $\Tr(\r \ps) = \Tr_{\CC^s} (\ph) (\ps \ox \1)$ for
$\ph \in_R \cP(A \ox B \ox \CC^s)$,
Lemma \ref{lem:concentrationTr} yields
\begin{equation}
\Pr\left\{ \left| \Tr\rho\ps - \frac{1}{d_Ad_B} \right| >
\frac{\d}{d_Ad_B} \right\}
  \leq 2 \exp\left( -s\frac{\d^2}{6\ln 2} \right) \,.
    \label{eq:expBoundDistill}
\end{equation}
By the triangle inequality and the union bound over a
$\frac{\d}{d_Ad_B}$-net $\cN$ for Schmidt-rank two states in $A\ox B$
(Lemma~\ref{lem:netk}), with
$|\cN| \leq \left({10 d_A d_B}/{\d}\right)^{4(d_A+d_B)}$,
we obtain, putting $\d=1/10$ and using $d_B \geq d_A$,
\begin{equation*}
\Pr\left\{ \exists\ps\text{ of Schmidt-rank }2\
  \left| \Tr\rho\ps - \frac{1}{d_Ad_B} \right| > \frac{2\d}{d_Ad_B} \right\}
  \leq 2\left(\frac{10d_B^2}{\d}\right)^{8d_B}
    \exp\left( -s\frac{\d^2}{6\ln 2} \right).
\end{equation*}

If for all Schmidt-rank two states $\ph$, $\left| \Tr\rho\ps -
\frac{1}{d_Ad_B} \right| \leq \frac{2\d}{d_Ad_B}$, then for all rank two
projectors $P,Q$ and for any state $|\xi\>$ in the support of $P \ox Q$,
\begin{equation*}
\smfrac{1{-}2\d}{d_Ad_B} \; \<\xi| (P \ox Q) |\xi\> \leq
\<\xi| (P \ox Q) \rho (P \ox Q) |\xi\>
\leq \smfrac{1{+}2\d}{d_Ad_B} \; \<\xi| (P \ox Q) |\xi\> \,,
\end{equation*}
and therefore
\begin{equation*}
\smfrac{1{-}2\d}{d_Ad_B} \; P \ox Q \leq (P \ox Q) \rho (P \ox Q)
\leq \smfrac{1{+}2\d}{d_Ad_B} \; P \ox Q \,.
\end{equation*}
and $\smfrac{4(1{-}2\d)}{d_Ad_B} \leq \Tr\bigl( (P \ox Q) \rho \bigr)
\leq \smfrac{4(1{+}2\d)}{d_Ad_B}$.
Thus, for $\sigma$ defined in \eq{sig2}
\begin{equation*}
\frac{1-2\d}{1+2\d}\cdot\frac{1}{4} P \ox Q \leq \s
           \leq \frac{1+2\d}{1-2\d}\cdot\frac{1}{4}P \ox Q.
\end{equation*}
The choice $\d \leq 1/10$ will secure that $\|\s-\frac{1}{4}P \ox Q\|_\infty
\leq 1/8$ and we are done.
\end{proof}

\begin{remark}
We began our study of correlation by considering the entanglement
of formation and mutual information for the maximally mixed state $\r_{AB}$
on a random subspace $S$ of dimension $s$, before moving on to
study mixed states with the measure induced by tracing over part of
a random pure state. To end, we note that both
Theorems \ref{thm:oneway-bounds} and \ref{thm:one-copy-undistillable}
apply unaltered if $\r_{AB}$ is chosen as a random maximally mixed state
instead of according
to $\cS_s(A\ox B)$. Not even the proofs need to change: the crucial
applications of Lemma \ref{lem:concentrationTr} in Eqs.~(\ref{eq:expBoundI})
and (\ref{eq:expBoundDistill}) give exactly the same estimates for the
new distribution.
\end{remark}

\section{Multiparty entanglement} \label{sec:multiparty}

With a little more work, and building upon the results obtained so far,
we can learn a good deal about the entanglement properties of generic
random multipartite states.
To that end, let $\ph \in_R {\cal P}\left( (\CC^d)^{\ox n} \right)$;
we could easily allow for different local dimensions but that would
only result in more cumbersome notation. Also,
let us label the $n$ subsystems by numbers $1,\ldots,n$.  A subset of
the parties is given the name $X\subset\{1,\ldots,n\}$, and its complement
$\overline{X}=\{1,\ldots,n\}\setminus X$.  Each $X$ thus defines a
bipartite cut, and we will freely call the cut $X$ as well.  Let
$\ph_X = \Tr_{\overline{X}} \ph$ denote the state reduced to the systems
in $X$.

\noindent The questions we address here are the following:
\begin{enumerate}
  \item Entropy of entanglement across any bipartite cut.
  \item Entanglement of formation and separability of reduced states
    on an arbitrary set of $k<n$ parties.
  \item Distillability of maximal entanglement between arbitrary
    pairs of parties by LOCC between all parties.
\end{enumerate}
For the first two questions, either
the local dimension $d$ or the number of parties $n$ can be treated as
the asymptotic parameter; the important thing, in fact, is that the combination
$d^n$ become large. For the distillability question, however, our results
will only be valid for large $d$.

\begin{corollary} \label{cor:multi-entanglement}
Let $\ph \in_R {\cal P}\left( (\CC^d)^{\ox n} \right)$ be a random
state, and $\a>0$. Then
\begin{equation}
  \Pr\bigl\{ \exists X\  E(\ph_{X,\overline{X}}) = S(\ph_X)
        < x\log d - \a - \b_X \bigr\}
         \leq 2^{n-1} \exp\left( -\frac{(d^n-1)C_3\a^2}{n^2(\log d)^2} \right),
  \label{eq:allcuts}
\end{equation}
where $x = \min(|X|,|\bar{X}|)$,
$C_3=(8\pi^2\ln 2)^{-1}$ is the same as in
Theorem \ref{thm:subsystemConvergence} and $\b_X=\frac{1}{\ln 2}d^{2x-n}$.
\end{corollary}
In other words, it is $d^n$-likely
that $\ph$ is highly entangled across any bipartite cut and
almost maximally entangled across any cut such that $|X|\neq n/2$.

\begin{proof}
This follows immediately from Theorem~\ref{thm:subsystemConvergence}
and the union bound on all $2^{n-1}$ cuts $X$ with $x\leq n/2$.  The
parameter $\beta_X$ is just $\beta$ in
Theorem~\ref{thm:subsystemConvergence} with the proper dimensions.

Note that we cover the case $d=2$, too, since there the Lipschitz
constant can be bounded by $\sqrt{8}\log 3$ and in
Eq.~(\ref{eq:allcuts}) we have substituted the much larger
$\sqrt{8}n\log d$.
\end{proof}

\begin{corollary} \label{cor:multi-EoF}
Let $\ph \in_R {\cal P}\left( (\CC^d)^{\ox n} \right)$ be a random
state, and consider arbitrary $X\subset\{1,\ldots,n\}$ of cardinality $x$
and arbitrary cuts within
$X$ into disjoint subsets, $X_{1},X_{2}$ of sizes $x_1 \leq x_2$.
Then there exist absolute numerical constants $M_1$ and $M_2$ such that
\begin{enumerate}
  \item If $x > n/2 + M_1 \frac{1}{\log d}\log\frac{n\log d}{\a}$,
        $0<\a<1$ and $\b=\frac{1}{2\ln 2}\,d^{x_1-x_2}$,
        it is $d^x$-likely that for all $X_1$,
        $E_f(\ph_{X}) \geq x_1 \log d - \a - \b$.

  \item If $x < n/2 - M_2 \frac{1}{\log d}\log\frac{n\log d}{\e^2}$
        and $\e>0$, it is $d^{n-x}$-likely
        that for all $X_1$, $E_f(\ph_X) \leq \e$.

  \item If $x < n/3 - \frac{1}{\log d}$,
        it is $d^{n-3x}$-likely that $\ph_X$ is separable (as a multiparty state
of $x$ parties).
\end{enumerate}
\end{corollary}
\begin{proof}
For each $X$, the claims are simply Parts 1-3 of Theorem
\ref{thm:random-EoF} applied to $\r$ with total dimension $d^x$
and with rank $d^{\,n-x}$.  The worst case $X$ is taken care of
by a union bound over all $X$ and all possible cuts, of which there
are at most $3^n$ in total.

Note that our proof of Part 3 in Theorem~\ref{thm:random-EoF}
actually shows separability for every decomposition of the system
into arbitrary subsystems, because it uses only the bound on the
purity and the result of~\cite{GB03} to that effect.
\end{proof}

\medskip
\begin{remark}
Observe that the thresholds for the group sizes become, for
fixed $n$ and $d\rightarrow\infty$, $n/2$, $n/2$ and $n/3$.
The findings of Corollary~\ref{cor:multi-EoF} should be compared
to numerical investigations reported in~\cite{KZM02}: there the
threshold $n/3$ was argued heuristically based on the knowledge
of the expectation of $\Tr\rho^2$ and the postulate that it
would exhibit measure concentration. Interestingly, the numerical
studies indicate that the reduced state already becomes PPT at
$x \sim n/2$.
\end{remark}

\begin{corollary} \label{cor:multi-dist}
With $n$ fixed, consider the following one-shot protocol for distilling
entanglement between an arbitrary pair chosen from among $n$ $d$-dimensional
systems:

\parbox{6in}{
Let $\{\ket{e_j}\}$ be an agreed-upon local basis for each party, and
let $X$ denote the chosen pair.  Each party $i \in \overline{X}$
measures in his local basis $\{\ket{e_j}\}$ and sends the
result $j_i$ to $X$.  Let
$\ph_{X,J}$ denote the resulting {\em pure} state in $X$, where $J$
is one of the $d^{n-2}$ possible measurement outcomes.
}

\noindent Then it is likely that $\forall X,J$, $E (\ph_{X,J}) \geq \log d -
\frac{1}{\ln 2} - \a$.
In other words, there is \emph{one} protocol which allows any pair of
parties to distill almost $\log d$ ebits between them.
\end{corollary}
\begin{proof}
Note that $\ph_{X,J} \in_R \cP(X)$.  The claim then follows from Theorem
\ref{thm:subsystemConvergence} and the union bound.
\end{proof}

\begin{remark}
The yield of  the above distillation protocol
is a nearly maximally entangled state between the members of the pair.
This feature of generic multiparty entanglement is also shared by
the cluster state of Briegel and Raussendorf~\cite{BR01}; in the language
of their paper, random multiparty pure states are likely to have maximally
persistent entanglement and to be maximally connected, modulo the fact
that the state distilled in Corollary \ref{cor:multi-dist}
is not \emph{exactly}
a maximally entangled state. Along the same lines, when $n$ is fixed
and $d$ is large, the protocol presented here can be used
to distill arbitrary pairwise entanglement, which in turn allows any arbitrary
pure state between the $n$ parties to be prepared. The efficiency, however,
could be very poor.
\end{remark}

\medskip\noindent
It is clear that there are innumerable other entanglement parameters
one could investigate for $\ph \in_R {\cal P}\left( (\CC^d)^{\ox n} \right)$.
The question of identifying the maximal yield for states other than
bipartite maximally entangled states seems to be particularly interesting
given the difficulty inherent in studying such questions for non-generic
states.

\section{Derandomization of superdense coding} \label{sec:superdense}

Superdense coding of quantum states was introduced in \cite{HHL03}: there it
was shown that, in the large-dimensional asymptotics, the state of two
qubits can be communicated exactly with high probability using one
ebit of entanglement and one transmitted qubit provided
the sender has full knowledge of the communicated state. (This is known
as the \emph{visible} scenario.) However, the
protocol in \cite{HHL03} also requires one shared bit of randomness
per two qubits communicated.

Theorem \ref{thm:subspace} suggests an alternative protocol that
does not require shared randomness:
Let the sender and receiver possess systems $B$ and $A$ initially.
(Note that this convention is opposite to common usage, but has the
advantage that $d_B \geq d_A$ in accord with the rest of this paper.)
Let $\a$, $\b$, and $\G$ be as defined in Theorem \ref{thm:subspace},
$d_B = d_A (\log d_A)^{2.5} \; \G^{-1} \a^{-2.5}$, so that a subspace $S$
as described in Corollary \ref{cor:subspace-fidelity}) can be chosen
with $s = d_A^2$.
Here, $\b<\a$, so that for every $\ket{\ph}\in S$ there exists a
maximally entangled state $\ket{\Phi}\in A\ox B$ with
$|\braket{\ph}{\Phi}|^2 \geq 1 - 2\sqrt{\a}$.
Starting from a fixed maximally entangled state $\Phi_0$ on $A\ox B$,
the sender can prepare any quantum state $\ket{\ph} \in S$
of $\log s = 2 \log d_A$
qubits in the receiver's laboratory by applying a unitary transformation
$U$ to $B$ such that $\ket{\Phi} = (\1\ox U)\ket{\Phi_0}$ and sends
his system to the receiver, who projects the state into the subspace
$S$ (and substitutes an arbitrary state if the projection fails).
It is evident that this protocol achieves what we aimed for.

\begin{theorem} \label{thm:superdense}
Asymptotically, $2 \log d_A$ qubits can be communicated visibly by
using $\log d_A$ ebits and $\log d_B = \log d_A + 2.5 \log \log d_A -
\log (\G \a^{2.5})$ qubits of communication.
The fidelity is $\geq 1 - 2\sqrt{\a}$.
\qed
\end{theorem}

Note a technical point, however\,: we pay a certain price for not
having to spend shared randomness. The protocol of \cite{HHL03} produces
an \emph{exact} copy of the target state when it succeeds, which occurs
with high probability. The protocol we propose here always succeeds, but
is not guaranteed to be exact. While the distinction is unimportant in
practice because the fidelity in our protocol can be made arbitrarily high,
the probabilistic-exact formulation is nonetheless the stronger criterion
from a theoretical point of view.
We do not know if the small sacrifice of fidelity
is essential for the derandomization or if, instead, a derandomized
probabilistic-exact protocol exists.

\section{Discussion} \label{sec:discussion}

%
%

We have seen that exponentially tight measure concentration, along
with careful attention to the achievable exponents, leads to many
interesting statements about the ubiquity, in composite systems, of subspaces
and states with rather extreme properties.
Specifically, many natural entanglement quantities
are amenable to techniques from the theory;
we found that there abound large subspaces containing only almost
maximally entangled states, whereas states supported on such subspaces
can be shown to yield almost no distillable correlation in the form of
entanglement, secret key or common randomness. In fact, in sharp contrast
to the difficulty one encounters for specific examples, our techniques
yield very good approximations to the values
of these correlation quantities for generic random states.

Figure \ref{fig:asymptotics} collects many of our results on correlation
measures. Perhaps its most striking feature is the gap between a random state's
entanglement of formation and its distillable correlation as the rank
of the random state approaches the total dimension. In that regime, the gap
is as large as it would be between a maximally entangled state and a product
state.
Thus, strong irreversibility of entanglement, quantified as a gap between
preparation cost and distillability, seems to be generic in large
systems. While the conclusion relies on the assumption that the
entanglement of formation is additive, the only way to evade it
would be for additivity to fail very drastically for random states.

We have also begun exploring the effects of measure concentration in
multipartite systems:
once again, the states seem to behave in quite unexpected, even
counterintuitive, but ultimately rather uniform ways.
Random pure states, for
example, almost always have near-maximal distillable entanglement between
any pair of parties, provided all other parties are allowed to participate
in the distillation protocol. For an $n$-party state, we can also identify
$x=n/2$ as the point at which the state of a subset of $x$ parties
transitions from having near maximal entanglement of formation to near-zero
entanglement of formation. Also, below $x=n/3$, the state becomes
separable, confirming numerical evidence and heuristic reasoning
from~\cite{KZM02}.

The large subspace of almost maximally entangled states mentioned earlier
also has
a constructive consequence: it allows us to get rid of the
shared randomness in previous protocols for ``superdense coding
of quantum states''. The result presented here, moreover, can be considerably
strengthened: optimal protocols for superdense coding of entangled
quantum states are presented in~\cite{AHSW04}.


Our work leaves open a number of questions, many of which we've
mentioned in the course of our presentation. We collect here
some of those we find most interesting:
\begin{enumerate}
\item
There are some precedents in the literature for our results on
entangled subspaces. If one relaxes the condition on the subspace,
asking only that it contain no product states, as opposed to exclusively
maximally entangled states, then the dimension of $S$ can be improved;
Parthasarathy recently demonstrated that $S$ could be taken to be of dimension
$d_A d_B - d_A - d_B + 1$, and that this is maximal~\cite{P04}.
For the sake of comparison, by taking $\a = \smfrac{1}{2} \log d_A$
in Theorem \ref{thm:subspace}, we find that there exists a
subspace $S$ of dimension
$\lfloor \smfrac{d_A d_B}{9917} \rfloor$,
all of whose states have entanglement at least $\smfrac{1}{2}\log d_A - \b$.
While the gap between the two results is small if measured in qubits,
it is still significant
in absolute terms. It is, therefore, natural to ask
how entangled the states of $S$ can be while still attaining
Parthasarathy's bound.
\item
The techniques used here are inadequate for exploring the transition from
near-maximal to near-zero entanglement of formation in rank-$s$ random
states. What is the behavior of $E_f$ in the transition region?
\item
How much can be said about the additivity conjecture for random
quantum states? The results in this paper, for example, can be used to show
that a random pure state of a sufficiently high-dimensional four-party
system will
not violate the superadditivity conjecture for the entanglement of
formation. That, however, is insufficient to rescue us from the
conditional nature of our conclusions about $E_c$ and $E_d^\rar$ based
on results for $E_f$ and $I^\rar$.
\item
Theorems \ref{thm:oneway-bounds} and \ref{thm:one-copy-undistillable} can be
interpreted as evidence that, when $s \gg d_B \log d_B$,
rank-$s$ random states on $A\ox B$ are actually undistillable. This
would be very interesting to decide, as these same random
states are likely to have near-maximal entanglement of
formation, so being simultaneously undistillable would make them
extreme examples of bound entanglement.
\item
While we have studied the mixed-state measures induced by taking the partial
trace over
a larger system, there are other proposals for measures on the set of
mixed states. Are our results still true, for example, if one substitutes
the Bures measure~\cite{Hall98,SZ04} for our choice?
\item
How does one construct random states? Are there physical processes that
will naturally produce states of the type we have studied here?
One possibility for engineering them would be to use the pseudorandom
unitaries of \cite{EWSLC03}. To what extent will the deviation from
the true Haar measure affect our conclusions~\cite{WH04}?
\end{enumerate}

There is no question that \emph{random} entangled states are far easier
to understand than \emph{all} entangled states. While here we have
focussed primarily on entanglement measures, it could even be the case that
the theory of interconversions undergoes a similar drastic simplification.
Perhaps equivalence via LOCC for random states can be completely
resolved, up to the inevitable exceptional sets,
a speculative note on which we would like
to end this paper.

\acknowledgments

Thanks to Noah Linden for interesting conversations on multiparty
entangled states and for comments on an earlier
version of our manuscript.
A large part of this work was done during a visit of AW's to
the Institute for Quantum Information in spring 2004.
AW gratefully acknowledges the hospitality of IQI during this time;
he furthermore acknowledges support from the EU under European Commission
project RESQ, contract No.~IST-2001-37559.
DL and PH are supported by National Science Foundation Grant
No.~EIA-0086038. PH acknowledges support from the Sherman Fairchild
Foundation.

\bibliographystyle{plain}
\bibliography{subspace}

\begin{thebibliography}{10}

\bibitem{AHSW04}
A.~Abeyesinghe, P.~Hayden, G.~Smith, and A.~Winter.
\newblock Optimal superdense coding of entangled states.
\newblock ar{X}iv:quant-ph/0407061.

\bibitem{AC98}
R.~Ahlswede and I.~Csisz\'{a}r.
\newblock Common randomness in information theory and cryptography - {P}art
  {II}: {CR}-capacity.
\newblock {\em {IEEE} Trans. Inf. Theory}, 44:225--240, 1998.

\bibitem{BDSW96}
C.~H. Bennett, D.~P. DiVincenzo, J.~A. Smolin, and W.~K. Wootters.
\newblock Mixed state entanglement and quantum error correction.
\newblock {\em Phys. Rev. A}, 54:3824--3851, 1996.
\newblock ar{X}iv:quant-ph/9604024.

\bibitem{Braunstein96}
S.~L. Braunstein.
\newblock Geometry of quantum inference.
\newblock {\em Phys. Lett. A}, 247:169, 1996.

\bibitem{BR01}
H.J. Briegel and R.~Raussendorf.
\newblock Persistent entanglement in arrays of interacting particles.
\newblock {\em Phys. Rev. Lett.}, 86(5):910--913, 2001.
\newblock ar{X}iv:quant-ph/0004051.

\bibitem{CW03}
M.~Christandl and A.~Winter.
\newblock Squashed entanglement -- {A}n additive entanglement measure.
\newblock {\em J. Math. Phys.}, 45(3):829--840, 2004.
\newblock ar{X}iv:quant-ph/0308088.

\bibitem{DZ93}
A.~Dembo and O.~Zeitouni.
\newblock {\em Large {D}eviations {T}echniques and {A}pplications}.
\newblock Springer-Verlag, New York, 1993.

\bibitem{D03}
I.~Devetak.
\newblock The private classical capacity and quantum capacity of a quantum
  channel.
\newblock ar{X}iv:quant-ph/0304127.

\bibitem{DW03b}
I.~Devetak and A.~Winter.
\newblock Distillation of secret key and entanglement from quantum states.
\newblock ar{X}iv:quant-ph/0306078.

\bibitem{DW03a}
I.~Devetak and A.~Winter.
\newblock Distilling common randomness from bipartite quantum states.
\newblock ar{X}iv:quant-ph/0304196.

\bibitem{DW03c}
I.~Devetak and A.~Winter.
\newblock Relating quantum privacy and quantum coherence: an operational
  approach.
\newblock ar{X}iv:quant-ph/0307053.

\bibitem{DHO01}
M.~Donald, M.~Horodecki, and O.~Rudolph.
\newblock The uniqueness theorem for entanglement measures.
\newblock {\em J. Math. Phys.}, 43:4252--4272, 2002.

\bibitem{DP99}
J.~J. Duistermaat and J.~A.~C. Polk.
\newblock {\em Lie Groups}.
\newblock Springer-Verlag, Berlin, 1999.

\bibitem{EWSLC03}
J.~Emerson, Y.~S. Weinstein, M.~Saraceno, S.~Lloyd, and D.~G. Cory.
\newblock Pseudo-random unitary operators for quantum information processing.
\newblock {\em Science}, 302:2098, 2003.

\bibitem{FK94}
S.~K. Foong and S.~Kanno.
\newblock Proof of {P}age's conjecture on the average entropy of a subsystem.
\newblock {\em Phys. Rev. Lett.}, 72:1148--1151, 1994.

\bibitem{Fuchsv99}
C.~A. Fuchs and J.~van~de Graaf.
\newblock Cryptographic distinguishability measures for quantum mechanical
  states.
\newblock {\em IEEE Trans. Inf. Theory}, 45:1216--1227, 1999.

\bibitem{GB03}
L.~Gurvits and H.~Barnum.
\newblock Separable balls around the maximally mixed multipartite quantum
  states.
\newblock {\em Phys. Rev. A}, 68:042312, 2003.
\newblock ar{X}iv:quant-ph/0302102.

\bibitem{Hall98}
M.~J.~W. Hall.
\newblock Random quantum correlations and density operator distributions.
\newblock {\em Phys. Lett. A}, 242:123--129, 1998.

\bibitem{Harris92}
J.~Harris.
\newblock {\em {A}lgebraic {G}eometry: {A} {F}irst {C}ourse}, volume 133 of
  {\em {G}raduate {T}exts in {M}athematics}.
\newblock {S}pringer {V}erlag, 1992.

\bibitem{HHL03}
A.~Harrow, P.~Hayden, and D.~W. Leung.
\newblock Superdense coding of quantum states.
\newblock {\em Phys. Rev. Lett.}, 92:187901, 2004.
\newblock ar{X}iv:quant-ph/0307221.

\bibitem{HHT00}
P.~Hayden, M.~Horodecki, and B.~M. Terhal.
\newblock The asymptotic entanglement cost of preparing a quantum state.
\newblock {\em J. Phys. A}, 34(35):6891--6898, 2001.
\newblock ar{X}iv:quant-ph/0008134.

\bibitem{HLSW03}
P.~Hayden, D.~W. Leung, P.~W. Shor, and A.~Winter.
\newblock Randomizing quantum states: Constructions and applications.
\newblock arXiv:quant-ph/0307104, 2003.
\newblock Commun. Math. Phys., to appear.

\bibitem{HLS04}
P.~Hayden, D.~W. Leung, and G.~Smith.
\newblock Multiparty data hiding of quantum information.
\newblock In preparation.

\bibitem{H73}
A.~S. Holevo.
\newblock Bounds for the quantity of information transmittable by a quantum
  communications channel.
\newblock {\em Problemy pereda\v{c}i Informacii}, 9(3):3--11, 1973.
\newblock English translation: A.~S. Holevo, \emph{Probl. Inf. Transm.}
  9:177-183, 1973.

\bibitem{HHHO04}
K.~Horodecki, M.~Horodecki, P.~Horodecki, and J.~Oppenheim.
\newblock Locking entanglement measures with a single qubit.
\newblock ar{X}iv:quant-ph/0404096.

\bibitem{MH01}
M.~Horodecki.
\newblock Entanglement measures.
\newblock {\em Quantum Inf. Comp.}, 1(1):3--26, 2001.

\bibitem{HHH97}
M.~Horodecki, P.~Horodecki, and R.~Horodecki.
\newblock Inseparable two spin-$1/2$ density matrices can be distilled to a
  singlet form.
\newblock {\em Phys. Rev. Lett.}, 78:574, 1997.

\bibitem{HHH98}
M.~Horodecki, P.~Horodecki, and R.~Horodecki.
\newblock Mixed-state entanglement and distillation: {I}s there a ``bound''
  entanglement in nature?
\newblock {\em Phys. Rev. Lett.}, 80:5239, 1998.
\newblock ar{X}iv:quant-ph/9801069.

\bibitem{Jozsa94}
R.~Jozsa.
\newblock Fidelity for mixed quantum states.
\newblock {\em J. Mod. Opt.}, 41:2315--2323, 1994.

\bibitem{KZM02}
V.~Kendon, K.~{\.{Z}}yczkowski, and W.~Munro.
\newblock Bounds on entanglement in qudit subsystems.
\newblock {\em Phys. Rev. A.}, 66:062310, 2002.
\newblock ar{X}iv:quant-ph/0203037.

\bibitem{KW04}
M.~Koashi and A.~Winter.
\newblock Monogamy of quantum entanglement and other correlations.
\newblock {\em Phys. Rev. A}, 69:022309, 2004.
\newblock ar{X}iv:quant-ph/0310037.

\bibitem{L01}
M.~Ledoux.
\newblock {\em The concentration of measure phenomenon}, volume~89 of {\em
  Mathematical Surveys and Monographs}.
\newblock American Mathematical Society, 2001.

\bibitem{LP88}
S.~Lloyd and H.~Pagels.
\newblock Complexity as thermodynamic depth.
\newblock {\em Ann. Phys.}, 188(1):186--213, 1988.

\bibitem{L78}
E.~Lubkin.
\newblock Entropy of an $n$-system from its correlation with a $k$-reservoir.
\newblock {\em J. Math. Phys.}, 19:1028, 1978.

\bibitem{MS86}
V.~D. Milman and G.~Schechtman.
\newblock {\em Asymptotic theory of finite dimensional normed spaces}, volume
  1200 of {\em Lecture Notes in Mathematics}.
\newblock Springer-Verlag, 1986.

\bibitem{Nielsen00bk}
M.~A. Nielsen and I.~L. Chuang.
\newblock {\em Quantum computation and quantum information}.
\newblock Cambridge University Press, Cambridge, U.K., 2000.

\bibitem{OhyaP93}
M.~Ohya and D.~Petz.
\newblock {\em Quantum entropy and its use}.
\newblock Texts and monographs in physics. Springer-Verlag, Berlin, 1993.

\bibitem{OHHHH03}
J.~Oppenheim, K.~Horodecki, M.~Horodecki, P.~Horodecki, and R.~Horodecki.
\newblock A new type of complementarity between quantum and classical
  information.
\newblock {\em Phys. Rev. A}, 68:022307, 2003.
\newblock ar{X}iv:quant-ph/0207025.

\bibitem{P93}
D.~N. Page.
\newblock Average entropy of a subsystem.
\newblock {\em Phys. Rev. Lett.}, 71:1291, 1993.

\bibitem{P04}
K.~R. Parthasarathy.
\newblock On the maximal dimension of a completely entangled subspace for
  finite level quantum systems.
\newblock ar{X}iv:quant-ph/0405077.

\bibitem{P96}
A.~Peres.
\newblock Separability criterion for density matrices.
\newblock {\em Phys. Rev. Lett.}, 77:1413, 1996.
\newblock ar{X}iv:quant-ph/9604005.

\bibitem{S-R95}
J.~Sanchez-Ruiz.
\newblock Simple proof of {P}age's conjecture on the average entropy of a
  subsystem.
\newblock {\em Phys. Rev. E}, 52:5653, 1995.

\bibitem{SW01}
B.~Schumacher and M.~D. Westmoreland.
\newblock Approximate quantum error correction, 2001.
\newblock ar{X}iv:quant-ph/0112106.

\bibitem{S96}
S.~Sen.
\newblock Average entropy of a quantum subsystem.
\newblock {\em Phys. Rev. Lett.}, 77(1):1--3, 1996.

\bibitem{S04}
P.~W. Shor.
\newblock Equivalence of additivity questions in quantum information theory.
\newblock {\em Commun. Math. Phys.}, 246(3):453--472, 2004.
\newblock ar{X}iv:quant-ph/0305035.

\bibitem{SZ04}
H.-J. Sommers and K.~{\.{Z}}yczkowski.
\newblock Statistical properties of random density matrices.
\newblock ar{X}iv:quant-ph/0405031.

\bibitem{S03}
S.~Szarek.
\newblock The volume of separable states is super-doubly-exponentially small.
\newblock ar{X}iv:quant-ph/0310061.

\bibitem{Uhlmann76}
A.~Uhlmann.
\newblock The `transition probability' in the state space of a {$^*$}-algebra.
\newblock {\em Rep. Math. Phys.}, 9:273, 1976.

\bibitem{U62}
H.~Umegaki.
\newblock Conditional expectations in an operator algebra {IV} (entropy and
  information).
\newblock {\em Kodai Math. Sem. Rep.}, 14:59--85, 1962.

\bibitem{VPRK97}
V.~Vedral, M.~B. Plenio, M.~A. Rippin, and P.~L. Knight.
\newblock Quantifying entanglement.
\newblock {\em Phys. Rev. Lett.}, 78(12):2275--2279, 1997.

\bibitem{VC01}
G.~Vidal and J.~I. Cirac.
\newblock Irreversibility in asymptotic manipulations of entanglement.
\newblock {\em Phys. Rev. Lett.}, 86:5803--5806, 2001.
\newblock ar{X}iv:quant-ph/0102036.

\bibitem{VDC02}
G.~Vidal, W.~D\"{u}r, and J.~I. Cirac.
\newblock Entanglement cost of mixed states.
\newblock {\em Phys. Rev. Lett.}, 89:027901, 2002.
\newblock ar{X}iv:quant-ph/0112131.

\bibitem{vN27}
J.~von Neumann.
\newblock Thermodynamik quantenmechanischer {G}esamtheiten.
\newblock {\em Nachr. der Gesellschaft der Wiss. G\"{o}tt.}, pages 273--291,
  1927.

\bibitem{WH04}
Y.~S. Weinstein and C.~S. Hellberg.
\newblock Matrix element randomness, entanglement, and quantum chaos.
\newblock ar{X}iv:quant-ph/0405053.

\bibitem{Y91}
R.~M. Young.
\newblock Euler's constant.
\newblock {\em Math. Gaz.}, 75:187--190, 1991.

\bibitem{ZZF00}
P.~Zanardi, C.~Zalka, and L.~Faoro.
\newblock On the entangling power of quantum evolutions.
\newblock {\em Phys. Rev. A}, 62:030301, 2000.
\newblock ar{X}iv:quant-ph/0005031.

\bibitem{Z99}
K.~{\.{Z}}yczkowski.
\newblock On the volume of the set of mixed entangled states {II}.
\newblock {\em Phys. Rev. A.}, 60:3496, 1999.
\newblock ar{X}iv:quant-ph/9902050.

\bibitem{ZHSL98}
K.~{\.{Z}}yczkowski, P.~Horodecki, A.~Sanpera, and M.~Lewenstein.
\newblock Volume of the set of separable states.
\newblock {\em Phys. Rev. A.}, 58:883--892, 1998.
\newblock ar{X}iv:quant-ph/9804024.

\bibitem{ZS01}
K.~{\.{Z}}yczkowski and H.-J. Sommers.
\newblock Induced measures in the space of mixed quantum states.
\newblock {\em J. Phys. A: Math. Gen.}, 34, 2001.
\newblock ar{X}iv:quant-ph/0012101.

\end{thebibliography}

\end{document}